\documentclass[journal=jpccck,manuscript=article]{achemso}
\usepackage[version=3]{mhchem} 
\usepackage{graphicx}
\usepackage{epstopdf}
\usepackage{multirow}
\usepackage{dcolumn}
\usepackage{bm}
\usepackage{subfigure}
\usepackage{color}
\usepackage{amsmath}
\usepackage{amssymb}
\usepackage{amsfonts}
\usepackage{amsthm}
\usepackage{xspace}
\usepackage{mciteplus}
\mciteErrorOnUnknownfalse

\author{S. C. Rakesh Roshan}
\affiliation{Rajiv Gandhi University of Knowledge Technologies, Basar, Telangana-504107, India}
\alsoaffiliation{Department of Physics, National Institute of Technology-Warangal, Telangana, India}
\email{roshan@rgukt.ac.in}
\author{N. Yedukondalu}
\email{nykondalu@gmail.com}
\affiliation{Department of Geosciences, Center for Materials by Design, and Institute for Advanced Computational Science, State University of New York, Stony Brook, New York 11794-2100, USA}
\author{Rajmohan Muthaiah}
\affiliation{School of Aerospace and Mechanical Engineering, University of Oklahoma, Norman, OK 73019, USA}
\author{K. Lavanya}
\affiliation{Jawaharlal Nehru Technological University (JNTU), Hyderabad, 500085, Telangana, India.}
\alsoaffiliation{Telangana Social Welfare Residential Degree College for Women (TSWRDC-W), Nizamabad, 503002, Telangana, India.}
\author{P. Anees}
\affiliation{Materials Physics Division, Indira Gandhi Centre for Atomic Research,
Kalpakkam 603102, Tamil Nadu, India} 

\author{R. Rakesh Kumar}
\affiliation{Department of Physics, Nanomaterials and Energy Harvesting Research Lab, National Institute of Technology-Warangal, Telangana, India} 

\author{T. Venkatappa Rao}
\affiliation{Department of Physics, Nanomaterials and Energy Harvesting Research Lab, National Institute of Technology-Warangal, Telangana, India} 

\author{Lars Ehm}
\affiliation{Department of Geosciences, Center for Materials by Design, and Institute for Advanced Computational Science, State University of New York, Stony Brook, New York 11794-2100, USA}
\author{John B. Parise}
\affiliation{Department of Geosciences, Center for Materials by Design, and Institute for Advanced Computational Science, State University of New York, Stony Brook, New York 11794-2100, USA}

\date{\today}

\title[An \textsf{achemso} demo]
{Anomalous Lattice Thermal Conductivity in Rocksalt IIA-VIA Compounds}
\begin{document}
\begin{abstract}
Materials with an intrinsic (ultra)low lattice thermal conductivity ($k_L$) are critically important for the development of efficient energy conversion devices. In the present work, we have investigated microscopic origins of low $k_L$ behavior in BaO, BaS and MgTe by exploring lattice dynamics and phonon transport of 16 iso-structural MX (Mg, Ca, Sr, Ba and X = O, S, Se and Te) compounds in the rocksalt (NaCl)-type structure by comparing their lattice transport properties with the champion thermoeletric iso-structural material, PbTe. Anomalous trends are observed for $k_L$ in MX compounds except the MgX series in contrast to the expected trend from their atomic mass. The underlying mechanisms for such low $k_L$ behavior in relatively low atomic mass systems namely BaO, BaS and MgTe compounds are thoroughly analyzed. We propose the following dominant factors that might be responsible for low $k_L$ behavior in these materials: 1) softening of transverse acoustic (TA) phonon modes despite low atomic mass, 2) low lying optic (LLO) phonon modes fall deep into acoustic mode region which enhances overlap between longitudinal acoustic (LA) and LLO phonon modes which increases scattering phase space, 3) short phonon lifetimes and high scattering rates, 4) relatively high density ($\rho$) and large Gr\"uneisen parameter. Moreover, tensile strain also causes a further reduction in $k_L$ for BaO, BaS and MgTe through phonon softening and near ferroelectric instability. Our comprehensive study on 16 binary MX compounds might provide a pathway for designing (ultra)low $k_L$ materials even with simple crystal systems through phonon engineering. 
\end{abstract}
\clearpage


\section{Introduction}
Discovery of materials with low lattice thermal conductivity gained tremendous interest due to their potential applications in thermoelectrics,\cite{AM-2021,TE-2020,nature-2021,AAAS-2021,TE-kanishka-2021} thermal barrier coatings\cite{TBC-2021,CLARKE200522,TBC-2019,ZHANG201718}, thermal insulation \cite{Ultralow-2021,LIU2020702} and thermal energy management \cite{TEM-2021}. Extensive efforts have been put forward by researchers in this direction to develop suitable materials for energy conversion applications during the last decade. Recently, binary alkaline-earth chalcogenides, MX (Mg, Ca, Sr, Ba and X = O, S, Se and Te) are being received considerable attention due to their potential applications in multifarious fields such as catalysis, microelectronics\cite{Zimmer1985}, opto-electronics (light emitting, laser and magneto optical devices)\cite{Pandey1991,Asano1978,Nakanishi1993} and thermoelectrics\cite{Rajput2019,Rajput2020,Rajput2021} despite their simple crystal structure. Bulk MX $\&$ their 2D counterparts,\cite{Zheng2015,Rajput2019,Rajput2020,Rajput2021} p-type PbTe $\&$ MTe nanocrystals,\cite{Biswas2011} CaTe-SnTe,\cite{Orabi2016} heavily doped SrTe with PbTe\cite{Tan2016,Kim2017} and BaTe-PbTe\cite{Lo2012} are found to have excellent thermoelectric figure of merit (zT) in the range of 0.5-1.32.\cite{Rajput2021,Tao2021} This shows the viability of these materials as power generating thermoelectric (TE) devices and are capable of converting waste heat into electricity. In general, the conversion efficiency is characterized by a dimensionless quantity $i.e.$, zT = S$\sigma^2$T/$k$, where $k$ = $k_e$+$k_L$; S, $\sigma$, T are the Seebeck coefficient, electrical conductivity, and absolute temperature, and $k_e$ and $k_L$ are the electronic and lattice thermal conductivities, respectively. The complex interdependence among these S, $\sigma$ and $k$ parameters becomes challenging to discover the high zT materials. Therefore, materials with an intrinsic ultralow $k$ (especially $k_L$) provide a pathway for discovering high zT materials without degrading its charge transport. The MX compounds are extensively studied from theoretical perspective, which mainly focused on exploring structural phase transitions\cite{Struct-Ukraine-2020,Zagorac2017,Struct-bayraki-2009, Struct-Hao-2010}, elastic \cite{Elastic-1,Struct-Hao-2010}, lattice dynamics\cite{Struct-bayraki-2009,lattice-2,Lattice-3-US,MgXT}, electronic structure  \cite{Zagorac2017,JApBaO-Electronic2006,electronic-3-Sajjad2014,TranBlaha-US,MgX-Theory-Tairi2017}, metallization\cite{Struct-Ukraine-2020,Metallic-US, CaTe-Nature,SrTe-US}, thermodynamic\cite{Thermodynamic-1,Thermo-2}, thermoelectric properties\cite{Rajput2019,Rajput2020,Rajput2021} at ambient and/or high pressure/temperature conditions. However, very limited studies have been dedicated towards understanding the phonon transport in MgSe,\cite{MgSe2021} MgTe,\cite{MgTe2021} CaX (X = O, S, Se and Te)\cite{Yang2020} and MTe (M = Mg, Ca, Sr, Ba and Pb)\cite{Xia2018} compounds. Therefore, a detailed and comparative study on phonon transport of MX compounds provide insights to achieve (ultra)low $k_L$ materials through phonon engineering, which is essential for discovery of high zT materials. In the present work, we shed more light on understanding lattice dynamics, phonon transport and mechanical properties of 16 MX compounds at ambient conditions. Interestingly, we observe anomalous trends in $k_L$ for CaX (CaS $\textgreater$ CaO $\textgreater$ CaSe $\textgreater$ CaTe), SrX (SrSe $\textgreater$ SrO $\textgreater$ SrS $\textgreater$ SrTe) and BaX (BaTe $\textgreater$ BaSe $\textgreater$ BaS $\textgreater$ BaO) series. Especially, the observed anomalous\cite{Seko2015} trend in BaX (partly in SrX and CaX\cite{Yang2020}) series is in contrast to the expected trend from their atomic mass. Overall, among 16 compounds, we found BaO, BaS and MgTe exhibit low $k_L$ behavior over the studied temperature range of 300-800 K despite their low atomic mass in rocksalt NaCl-type (B1) structure. The underlying mechanisms for such abnormal trends and low $k_L$ behavior are extensively discussed through the computed lattice dynamics, phonon lifetimes, scattering rates, phonon group velocities at 300 K and mechanical properties. We also have investigated the effect of tensile strain on lattice dynamics and phonon transport of BaO, BaS and MgTe compounds are discussed extensively.  

The rest of the paper is organized as follows: In the next section, we briefly describe the computational details, methodology, various parameters used to perform the computation and crystal structure. Results and discussions concerning anharmonic lattice dynamics, lattice thermal conductivity and mechanical properties of the 16 MX compounds. Finally, we propose important observations that will be helpful to achieve (ultra)low $k_L$ in general and in particular for the MX compounds and finally summarized major outcomes of the present study.  

\section{Computational details, methodology and crystal structure}
All the first principles calculations have been performed using Vienna Ab-initio Simulation Package (VASP)\cite{VASP1996} for MX (Mg, Ca, Sr, Ba and X = O, S, Se and Te) compounds. The exchange-correlation was treated with PBEsol functional and the electron-ion interactions with pseudo-potential based projected augmented wave (PAW) approach. We have considered 10 and 6 valence electrons for alkaline-earth metals (Mg, Ca, Sr and Ba) and chalcogens (O, S, Se and Te) as plane wave basis orbitals. A plane wave cutoff energy of 520 eV was used for plane wave basis set expansion and a spacing of 2$\pi$ $\times$ 0.024 $\AA^{-1}$ for k-mesh in the irreducible Brillouin zone. Since MX compounds are polar semiconductors, Born effective charges and dielectric constants are calculated using density functional perturbation theory to capture dipole-dipole interactions.


Lattice dynamics and thermal conductivity of MX compounds are calculated by considering harmonic (2$^{nd}$) and anharmonic (3$^{rd}$) inter atomic force constants (IFCs) using the temperature dependent effective potential (TDEP)\cite{TDEP_PRB_2011,TDEP_PRB1_2013,TDEP_PRB2_2013} method. In the present work, we have considered the expansion of inter atomic force constants (IFCs) up to 3$^{rd}$ order and the corresponding model Hamiltonian is given as follows: 
\begin{equation}
H = U_0 + \sum\limits_{i}\frac{p_i^2}{2m_i} + \frac{1}{2!}\sum\limits_{ij}\sum\limits_{\alpha\beta}\Phi_{ij}^{\alpha\beta}u_i^\alpha u_j^\beta + \frac{1}{3!}\sum\limits_{ijk}\sum\limits_{\alpha\beta\gamma}\psi_{ijk}^{\alpha\beta\gamma}u_i^\alpha u_j^\beta u_k^\gamma 
\end{equation}


where p$_i$ and u$_i$ are momentum and displacement of atom $i$, respectively. $\Phi_{ij}^{\alpha\beta}$ and $\psi_{ijk}^{\alpha\beta\gamma}$ are 2$^{nd}$ and 3$^{rd}$ order force constant matrices, respectively. To compute these harmonic (2$^{nd}$) and anharmonic (3$^{rd}$) inter atomic force constants (IFCs), we have performed ab initio molecular dynamics (AIMD) simulations as implemented in the VASP at 300 K. The AIMD calculations were run for 5000 MD steps with time-step of 1 fs ($i.e.,$ 5 ps) with 128 atoms (4$\times$4$\times$4) supercell for all the MX compounds. For 2$^{nd}$ and 3$^{rd}$ order IFCs, interactions up to 9$^{th}$ nearest neighbors were included to ensure the convergence of calculated lattice dynamics and phonon transport properties. The temperature was controlled with a Nos\'e-Hoover thermostat\cite{Hoover1986,Hoover1985}. The lattice thermal conductivity is calculated by iteratively solving the full Boltzmann transport equation (BTE), including three-phonon and isotope scatterings from the natural distribution on a 25$\times$25$\times$25 q-point grid.


The thermal conductivity tensor is given by 
\begin{equation}
k^{\alpha\beta} = \frac{1}{(2\pi)^3}\sum\limits_{s}\int dqC_\lambda v_{\lambda\alpha} v_{\lambda\beta}\tau_{\lambda\beta} 
\end{equation}
where C$_\lambda$ is the contribution per mode $\lambda$ = ($s,q$) to specific heat, $\alpha$ and $\beta$ are Cartesian components,  v$_\beta$ and $\tau_\beta$ are phonon velocity and scattering time, respectively. 

The scattering rates are calculated from a full inelastic phonon Boltzmann equation which is given by
\begin{equation}
k_BTv_\lambda . \nabla T\frac{\partial n_{0 \lambda}}{\partial T} = \sum\limits_{\lambda'\lambda''}\bigg[ P^+_{\lambda \lambda' \lambda''}(\Psi_{\lambda''} - \Psi_{\lambda'} - \Psi_\lambda) + \frac{1}{2} P^-_{\lambda \lambda' \lambda''} (\Psi_{\lambda''} + \Psi_{\lambda'} - \Psi_\lambda) \bigg ]
\end{equation}
The left-hand side represents the phonon difussion induced by the thermal gradient $\nabla$T and n$_{0 \lambda}$ is the equilibrium phonon distribution function. While the right-hand side corresponds to the collision term for three-phonon interactions. $v_\lambda$ is the phonon velocity in mode $\lambda$, P$^+_{\lambda \lambda' \lambda''}$ and P$^-_{\lambda \lambda' \lambda''}$ are three phonon scattering rates for absorption ($\lambda$ + $\lambda'$ $\rightarrow$ $\lambda''$) and emission ($\lambda$ $\rightarrow$ $\lambda'$ + $\lambda''$) processes, respectively.

\begin{figure}
\includegraphics[width=6.0in,height=3.8in]{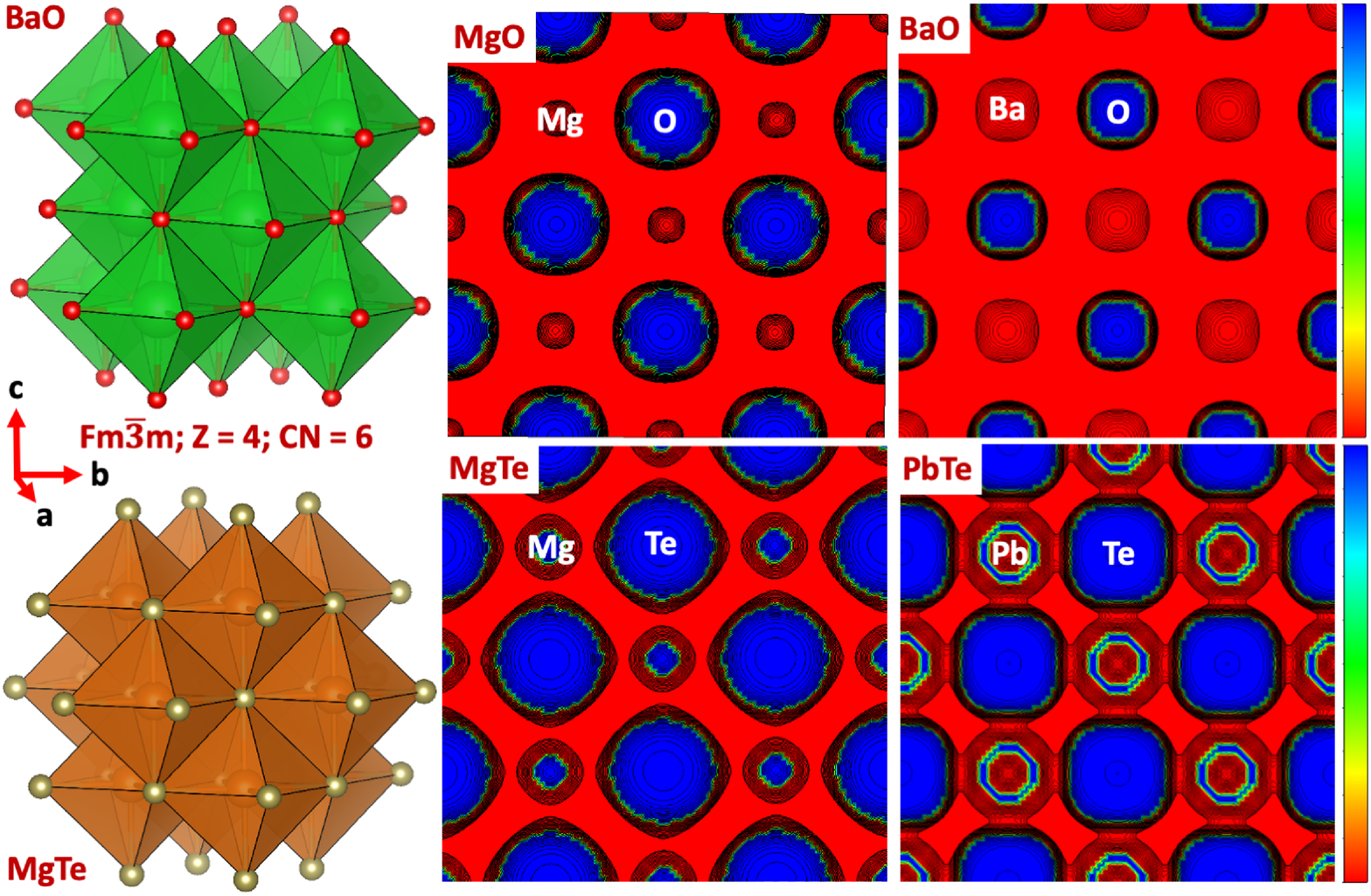}
\caption{Crystal structure of BaO and MgTe, electron localization function (ELF) of MgO, BaO, MgTe and PbTe along (001) plane.}
\label{fig:ELF}
\end{figure}

Binary alkaline-earth chalcogenides, MX (Mg, Ca, Sr, Ba and X = O, S, Se and Te) compounds except MgSe and MgTe crystallize in the face centred cubic (FCC) rocksalt NaCl (B1)-type structure (see Figure \ref{fig:ELF}) having space group $Fm\bar{3}m$ with Z = 4 formula units (f.u.) per unit cell at ambient conditions.\cite{CaX-Expt,CaTe-SrTe-EXPT,BaS-Expt,SrS-Expt,MgS-Expt,SrO-Expt,BaO-Expt} MgSe and MgTe exhibit rich polymorphism and they crystallize in rocksalt (B1), zinc-blende (B3), wurtzite (B4) and NiAs (B8)-type structures. The X-ray diffraction measurements reveal that MgTe crystallizes in B3\cite{MgTe-Wurzite-Waag1993} and B8\cite{MgTe-NiAs-Li1995} structures at ambient conditions. First principles calculations disclose that B3\cite{MgX-Theory-Tairi2017} phase for MgSe and both B3\cite{MgX-Theory-Tairi2017} and B8\cite{MgX-Structure-both-Duman2006,MgX-NaCl-Chakrabarti2000,MgSe-Te-Th-VanCamp1997}  for MgTe are thermodynamically stable structures at ambient conditions. Moreover, rocksalt-type B1 structure is dynamically stable (meta-stable) for both MgSe and MgTe compounds. Therefore, in the present work, we have considered B1 structure for all the MX compounds which allow us for direct comparison of the calculated properties among these 16 systems under investigation. Table \ref{table1} presents the calculated ground state equilibrium lattice constant for MX compounds in comparison with reported X-ray diffraction measurements\cite{MgO-Expt,MgS-Expt,CaO-Expt,MgSe-Expt,CaTe-SrTe-EXPT,CaX-Expt,SrO-Expt,SrS-Expt,BaS-Expt,PbTe-Expt-lattice-Bouad2003} and previous first principles calculations\cite{MO-Theory,MgX-Theory-Tairi2017,MgX-Mir2016,Rajput2019,Mg-CaX-Debnath2018,CaTe-Theory-CihanKrk2019,BaSe-Te-Drablia2017,PbTe-Zhang-Theory-a-2021,SrX-a-Theory,Rajput2021,PbTe-a-elastic-Xue2021} and there is a good agreement among them. In addition, we also calculated the electron localization function (ELF) for MgO, BaO, MgTe and PbTe compounds. As shown in Figure \ref{fig:ELF}, the metal cations (Mg$^{2+}$, Ba$^{2+}$) donate the electrons while anions (O$^{2-}$, Te$^{2-}$) gains the electrons indicating complete transfer of charge thus resulting in strong ionic bonding which is consistent with the fact that large electronegativity difference between Mg (1.31 in pauling scale), Ba (0.89) and O (3.44) results in strong ionic character for MgO (2.13) and BaO (2.55). The low electronegativity difference (0.79) between Mg (1.31) and Te (2.10) atoms suggest a polar covalent bonding along with ionic bonding in MgTe, while lone-pair of 6s$^2$ valence electrons is observed for PbTe along with sharing of electrons between Pb and Te. The distinct chemical bonding nature strongly influences $k_L$ of these materials. 



\section{Results and Discussion}
\subsection{Anharmonic lattice dynamics and thermal conductivity} 
\begin{figure}
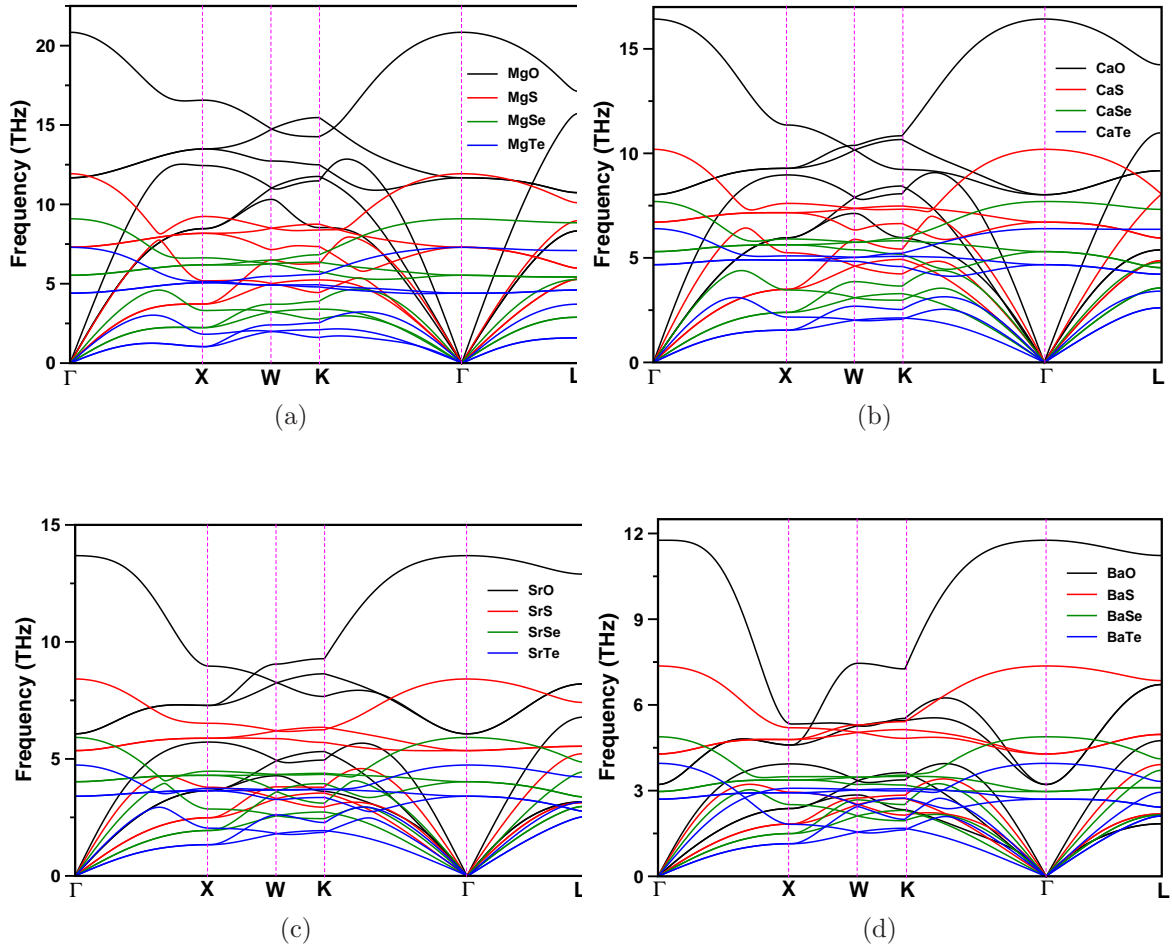

\centering
\subfigure[]{\includegraphics[width=3.0in,height=2.0in]{Figures/MgX-PD.eps}} 
\subfigure[]{\includegraphics[width=3.0in,height=2.0in]{Figures/CaX-PD.eps}} \vspace{0.3in} \\
\subfigure[]{\includegraphics[width=3.0in,height=2.0in]{Figures/SrX-PD.eps}} 
\subfigure[]{\includegraphics[width=3.0in,height=2.0in]{Figures/BaX-PD.eps}}
\caption{Calculated room temperature phonon dispersion curves of (a) MgX, (b) CaX, (c) SrX and (d) BaX compounds at PBEsol equilibrium volume; where X = O, S, Se and Te. Phonon softening is observed with increasing atomic mass of both alkaline-earth (Mg, Ca, Sr, Ba) and chaclogen (O, S, Se, Te) atoms.}
\label{fig:PD}
\end{figure}

\begin{figure}
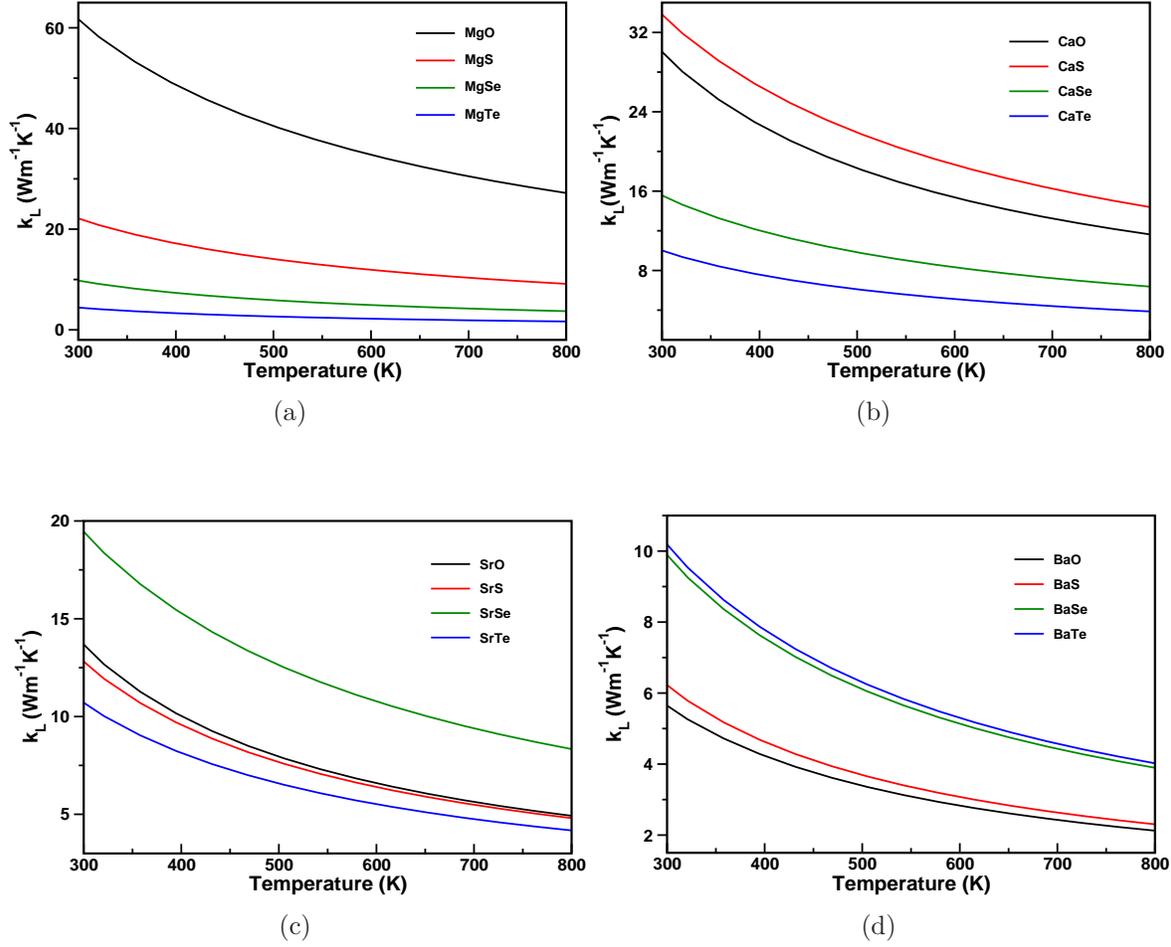

\centering
\subfigure[]{\includegraphics[width=3.0in,height=2.0in]{Figures/MgX-TC.eps}} 
\subfigure[]{\includegraphics[width=3.0in,height=2.0in]{Figures/CaX-TC.eps}} \vspace{0.3in} \\
\subfigure[]{\includegraphics[width=3.0in,height=2.0in]{Figures/SrX-TC.eps}} 
\subfigure[]{\includegraphics[width=3.0in,height=2.0in]{Figures/BaX-TC.eps}}
\caption{Calculated lattice thermal conductivity (k$_L$) of (a) MgX, (b) CaX, (c) SrX and (d) BaX compounds as a function of temperature; where X = O, S, Se and Te at PBEsol equilibrium volume. Anomalous trends are predicted for k$_L$ partly in CaX (CaS $\textgreater$ CaO $\textgreater$ CaSe $\textgreater$ CaTe), SrX (SrSe $\textgreater$ SrO $\textgreater$ SrS $\textgreater$ SrTe) and completely opposite trend for BaX (BaTe $\textgreater$ BaSe $\textgreater$ BaS $\textgreater$ BaO) compounds in contrast to the trends expected from their atomic mass.}
\label{fig:TC}
\end{figure}

Exploring lattice dynamics including anharmonic effects is crucial for understanding phonon transport in materials. As a first step, we have computed phonon dispersion curves (Figure \ref{fig:PD}) of MX compounds at 300 K and thoroughly analyzed them. As shown in Figure \ref{fig:PD}, no imaginary frequencies are found along high symmetry directions of the Brillouin zone indicating that all the investigated materials are dynamically stable. MX materials consist of 2 atoms per primitive cell resulting in (3N; N = number of atoms per primitive cell) 6 vibrational modes of which 3 are acoustic and 3 are optic modes. Dipole-dipole interactions are crucial for polar materials to describe phonon spectra correctly. These interactions are incorporated into dynamical matrix through calculated Born effective charges and high frequency dielectric constants which in turn produces a splitting between longitudinal optic (LO) and transverse optic (TO) phonon modes (Figure \ref{fig:PD}). Due to this LO-TO splitting, the three phonon optic modes split into two degenerate TO ($\omega_{TO}$) and one LO ($\omega_{LO}$) modes along $\Gamma$-direction. Large LO-TO splitting is observed in particular for MO compounds and it is increasing from MgO $\textless$ CaO $\textless$ SrO $\textless$ BaO, while it is decreasing from MO $\textgreater$ MS $\textgreater$ MSe $\textgreater$ MTe compounds (see Figure \ref{fig:PD} $\&$ Table S1). The MX compounds exhibit similar phonon band features and showed a significant phonon softening with increasing atomic mass from Mg $\rightarrow$ Ca $\rightarrow$ Sr $\rightarrow$ Ba and O $\rightarrow$ S $\rightarrow$ Se $\rightarrow$ Te and these features are consistent with previous first principles lattice dynamical calculations.\cite{Rajput2019,Rajput2020,Rajput2021,Xia2018} The  calculated lattice thermal conductivity ($k_L$) as function of temperature (300-800 K) is presented in Figure \ref{fig:TC}, the calculated $k_L$ values are decreasing with temperature for all the 16 compounds under investigation. Usually, materials with same crystal structure that consist of heavy elements, possess low $k_L$ compared to the ones with light elements due to their high atomic mass. As expected, $k_L$ is decreasing with increasing atomic mass of anion $i.e.,$ from MgO $\rightarrow$ MgS $\rightarrow$ MgSe $\rightarrow$ MgTe in MgX compounds, while $k_L$ shows anomalous trends for CaX (CaS $\textgreater$ CaO $\textgreater$ CaSe $\textgreater$ CaTe), SrX (SrSe $\textgreater$ SrO $\textgreater$ SrS $\textgreater$ SrTe) and BaX (BaTe $\textgreater$ BaSe $\textgreater$ BaS $\textgreater$ BaO) series of compounds. Especially, BaX series shows an opposite trend for $k_L$ in contrary to the expected trend from their atomic masses. However, the above trends are slightly altered, when $k_L$ is computed at the experimental lattice constant for SrX (SrSe $\textgreater$ SrO $\geq$ SrS $\textgreater$ SrTe) and BaX (BaSe $\geq$ BaTe $\textgreater$ BaO $\textgreater$ BaS) compounds but the overall trends remain more or less similar in both the cases (see Figures \ref{fig:TC} $\&$ S2). Usually, the PBE-GGA functional overestimates equilibrium lattice constant, therefore, the obtained k$_L$ values are underestimated when compared to the experimental values.\cite{Yang2020} Since, PBEsol functional underestimated the equilibrium lattice constant for PbTe (see Table \ref{table1}), the obtained $k_L$ values are overestimated compared to the ones obtained at the experimental lattice constant over the studied temperature range and this clearly demonstrates sensitivity of $k_L$ towards lattice constant(s) (see Figure S3). The predicted anomalous trends are originated from phonon softening observed from phonon dispersion curves (see Figure \ref{fig:PD}). As shown in Figure \ref{fig:PD}c $\&$ d, low lying TO modes of MO (M = Ca, Sr and Ba), SrS and BaS compounds fall deep into acoustic mode region in general and in particular for BaO\cite{Rajput2021} and SrO\cite{Rajput2021} compounds (see Figure \ref{fig:PD}) as observed for heavy metal binary PbTe.\cite{Xia2018} This enhances the overlap between TO and acoustic phonon modes which increases the scattering phase space, thus resulting in reduction of $k_L$ in those compounds. In addition, BaO shows extra phonon softening of acoustic modes near L point in the first Brillouin zone despite its low atomic mass compared to the rest of BaX (X = S, Se and Te) compounds. Among the 16 systems under investigation, interestingly, three compounds namely BaO, BaS and MgTe exhibit low $k_L$ $i.e.,$ below $\sim$ 6 Wm$^{-1}$K$^{-1}$ at 300 K. MgTe shows relatively low $k_L$ because of soft acoustic phonon modes near X and L points in the first Brillouin zone despite its low atomic mass compared to the rest of MTe (M = Ca, Sr, and Ba) compounds. This trend is consistent with the previous compressive sensing lattice dynamics (CSLD) study\cite{Xia2018} but in contrast to the phenomenological Debye-Callaway model study\cite{MTeRen2017} on lattice thermal transport of MTe (M = Mg, Sr, Ba and Pb) compounds. In addition, the $k_L$ values are severely underestimated using Debye-Callaway model (see Table \ref{table2}).  Overall, we observed two important aspects from phonon dispersion curves, which might be responsible for low $k_L$ behavior in BaO, BaS and MgTe compounds: 1) low lying TO phonon modes fall deep into acoustic mode region for BaO and BaS, 2) soft acoustic (TA) phonon modes of MgTe despite its low atomic mass over other heavy metal MTe (M = Ca, Sr, Ba) tellurides.

To further explore the underlying mechanisms which are responsible for the observed anomalous behavior of $k_L$ in MgTe, CaX, SrX and BaX compounds (see Figures \ref{fig:TC}, S1 and S2), we have calculated phonon mean free paths (MFPs), phonon lifetimes, group velocities and scattering rates. The calculated phonon MFPs as a function of frequency are presented in Figure S4. For all the MX compounds, a large portion of the phonon MFPs fall above the minimum inter-atomic distance or so called Ioffe-Regel limit. Therefore, phonon Boltzmann transport theory is good enough to describe thermal transport in MX compounds. In crystalline materials, the heat transport can be understood as the propagation of phonons and their scatterings among themselves. Since $k_L$ $\propto$ $\tau(\omega)$ and $v(\omega)$, therefore, materials with low $\tau(\omega)$ and $v(\omega)$ expected to have low $k_L$. 

\begin{figure}
\centering
\subfigure[]{\includegraphics[width=3.0in,height=2.0in]{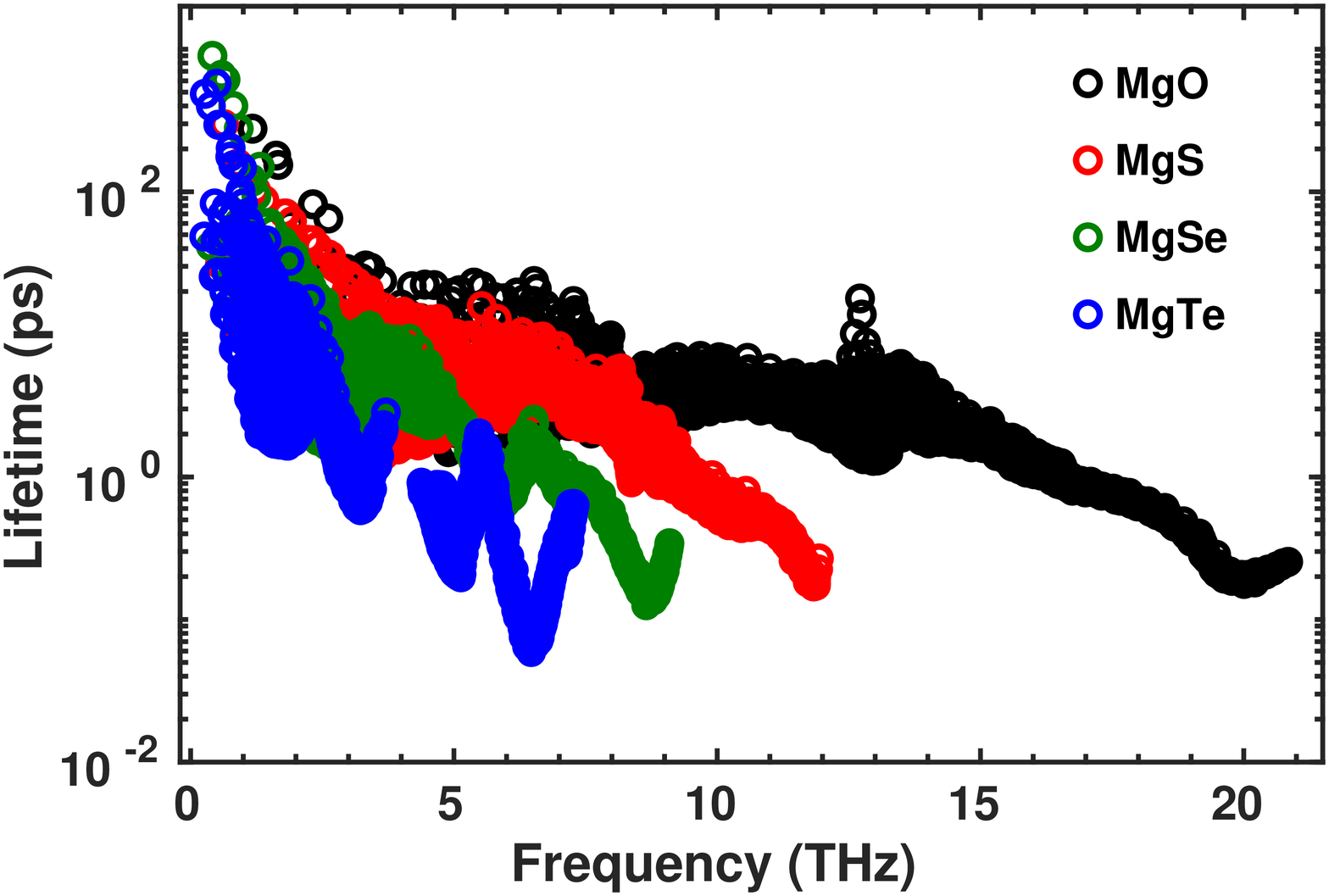}} 
\subfigure[]{\includegraphics[width=3.0in,height=2.0in]{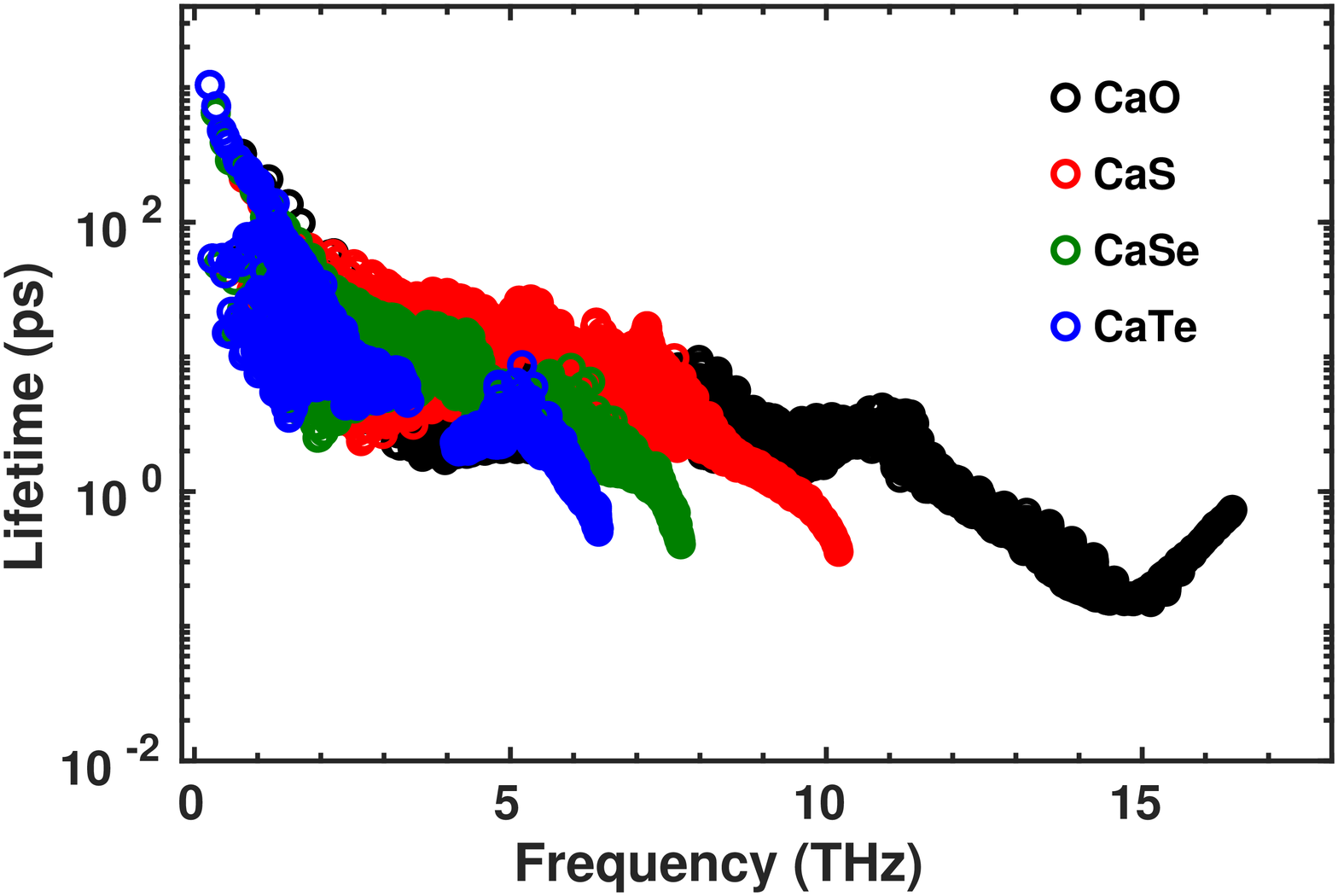}} \vspace{0.3in} \\
\subfigure[]{\includegraphics[width=3.0in,height=2.0in]{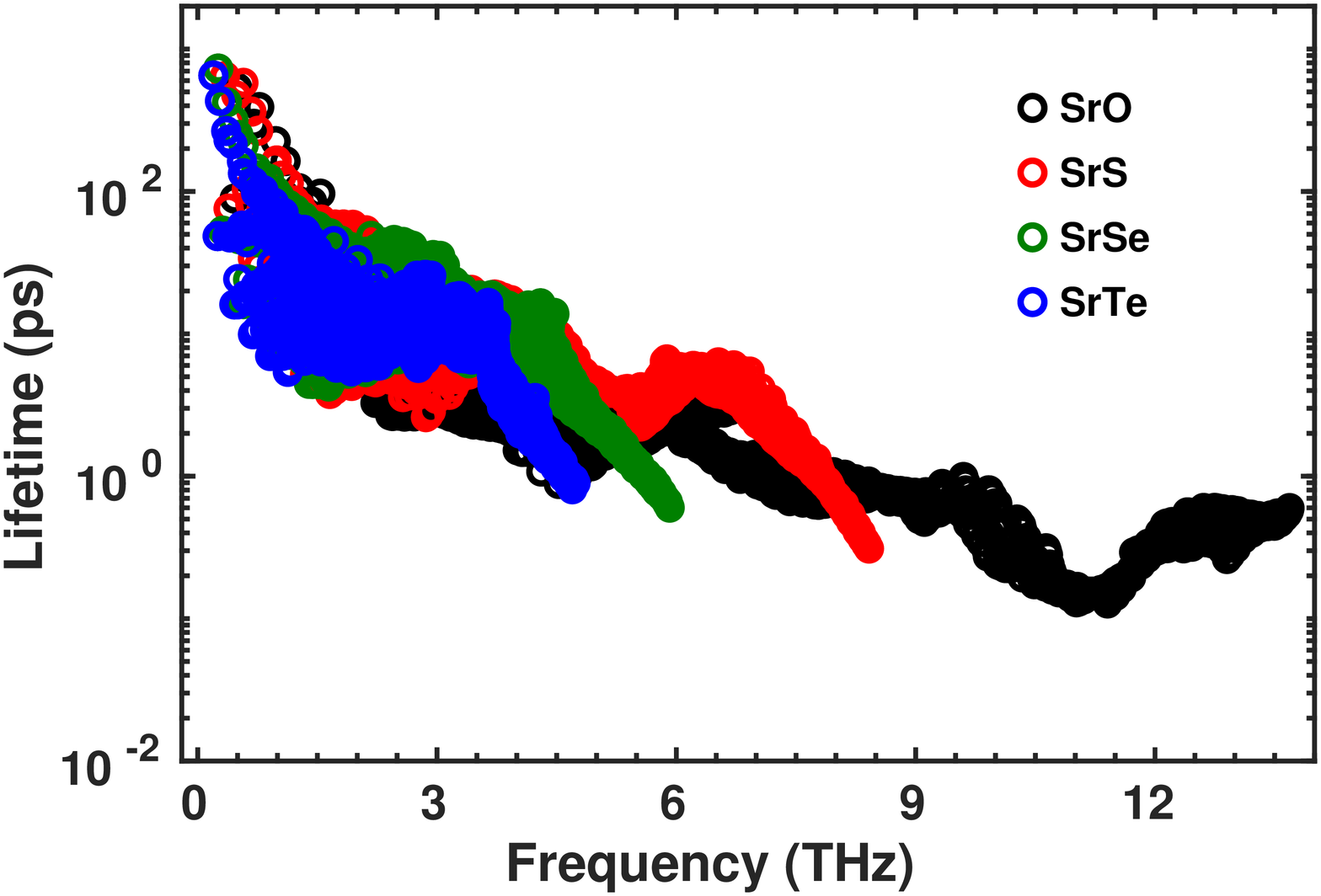}} 
\subfigure[]{\includegraphics[width=3.0in,height=2.0in]{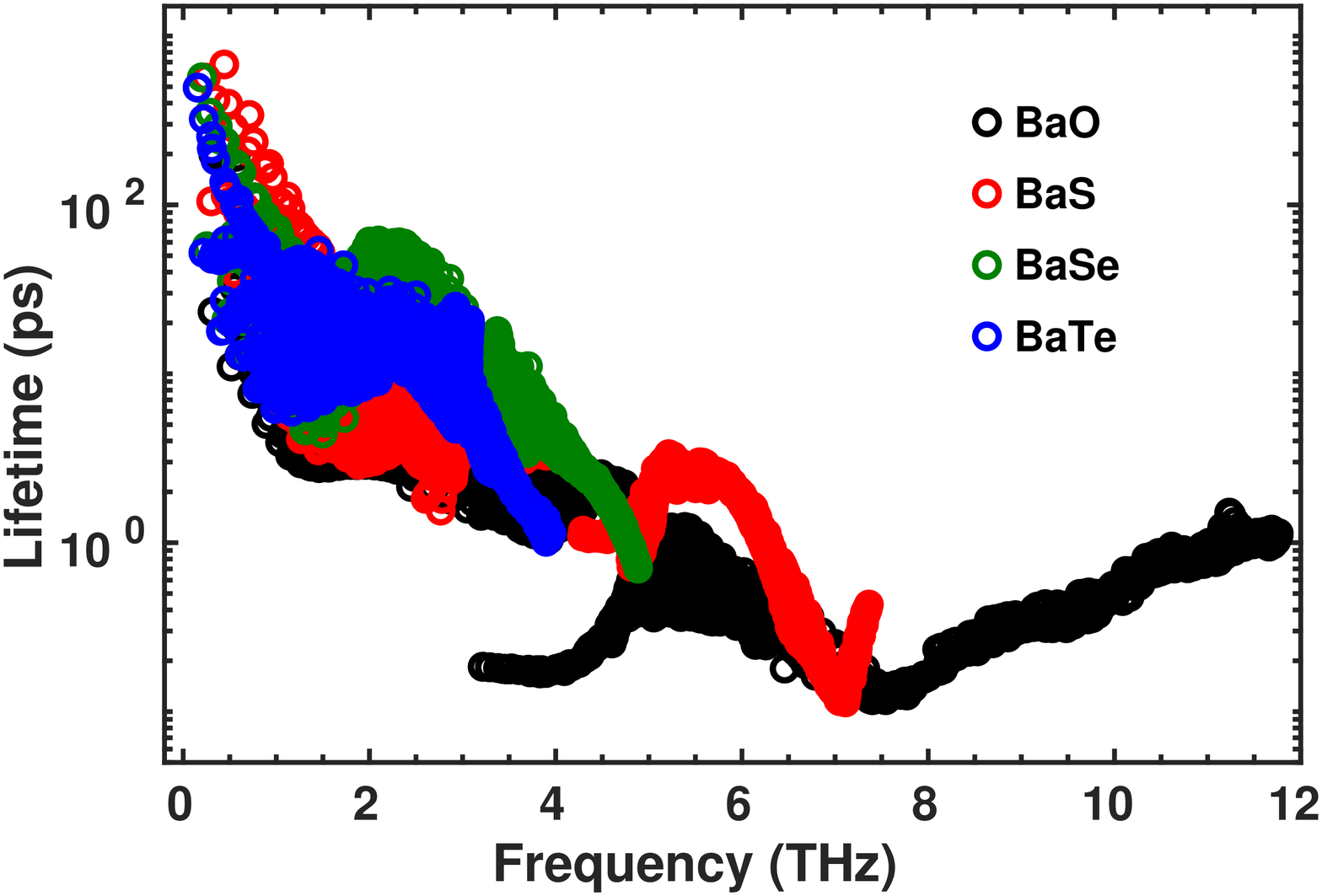}}
\caption{Calculated phonon lifetimes of (a) MgX, (b) CaX, (c) SrX and (d) BaX compounds as a function of frequency; where X = O, S, Se and Te.}
\label{fig:LT}
\end{figure}

As illustrated in Figure \ref{fig:LT}, phonon lifetime decreases from MgO $\textgreater$ MgS $\textgreater$ MgSe $\textgreater$ MgTe over the entire frequency range and the same trend is followed for $k_L$ in MgX. While CaO has relatively shorter phonon lifetimes than CaS in the frequency range of $\sim$ 2-8 THz (see Figure \ref{fig:LT}b), which might be reason for low $k_L$ of CaO thus results in the anomalous trend (CaS $\textgreater$ CaO $\textgreater$ CaSe $\textgreater$ CaTe) for $k_L$ in CaX series. This trend is consistent with the previous lattice thermal conductivity study\cite{Yang2020} on CaX compounds using ShengBTE. SrSe and SrO possess relatively highest and shorter phonon lifetimes in the frequency range of $\sim$ 1-4 THz and $\sim$ 2-4 THz, respectively. This could be a possible reason for the anomalous trend (SrSe $\textgreater$ SrO $\textgreater$ SrS $\textgreater$ SrTe) observed in SrX series for $k_L$. Finally, BaO and BaS have the shortest phonon lifetimes over BaSe and BaTe in turn they have low $k_L$ and this is consistent with the trend BaTe $\textgreater$ BaSe $\textgreater$ BaS $\textgreater$ BaO predicted for $k_L$ in BaX compounds.\cite{Seko2015} 

As shown in Figure S5, total scattering rates are obtained through summation of absorption, emission and isotope scattering rates from the three phonon processes for all the 16 MX compounds. The absorption scattering rates are largely dominated in the low frequency region (for instance, below 3 THz for BaO). In the low frequency region, phonon scattering processes probably occur through conversion of low energy phonon to a high energy phonon with an absorption of a phonon. The contribution of emission scattering rates gradually increase with frequency and are largely dominated at high frequency region, where phonon scattering processes probably occur through conversion of high energy phonon to a low energy phonon with an emission of a phonon. Finally, a moderate contribution from isotope scattering rates are observed over the entire frequency range, for instance, BaX compounds (see Figure S5).

\subsection{Effect of tensile strain on lattice thermal conductivity}
\begin{figure}
\centering
\subfigure[]{\includegraphics[width=3.0in,height=2.0in]{Figures/LMX-PD.eps}}
\subfigure[]{\includegraphics[width=3.1in,height=2.1in]{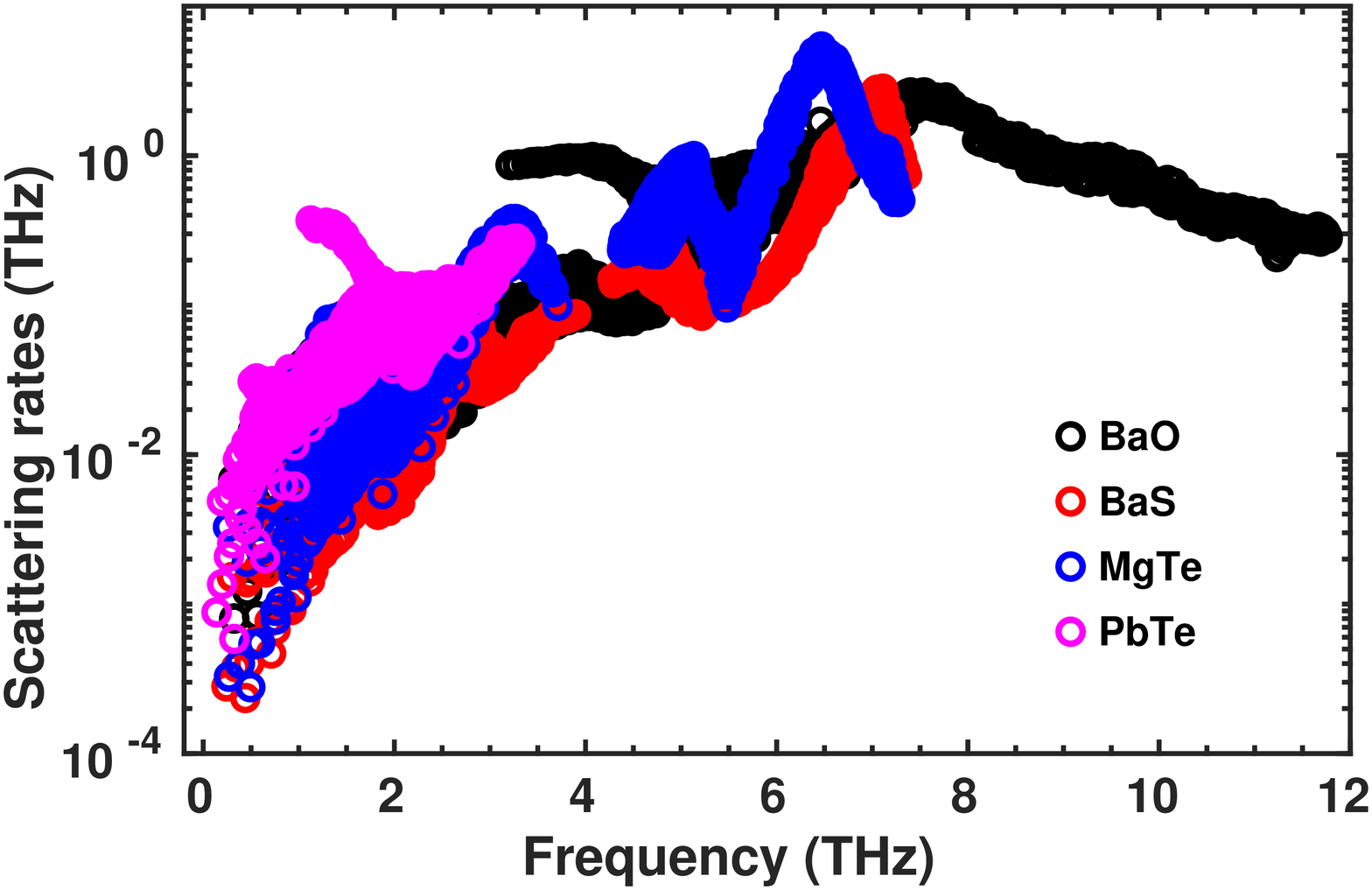}} \vspace{0.3in} \\
\subfigure[]{\includegraphics[width=3.1in,height=2.1in]{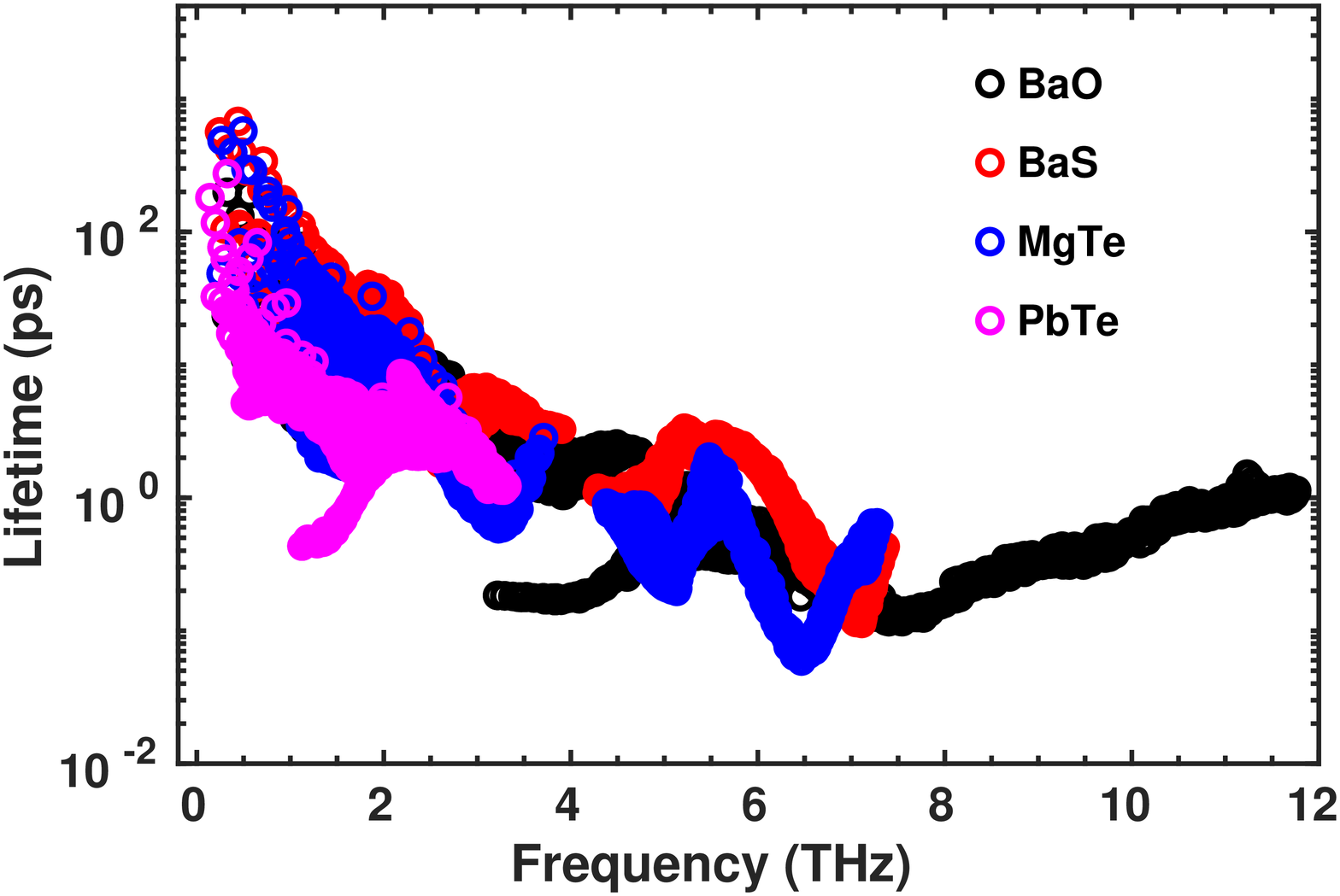}}
\subfigure[]{\includegraphics[width=3.0in,height=2.0in]{Figures/LMX-TC.eps}}
\caption{Calculated (a) phonon dispersion curves (b) phonon scattering rates (c) phonon lifetime, and (d) lattice thermal conductivity ($k_L$) of BaO, BaS, MgTe and PbTe compounds at PBEsol equilibrium volume. The obtained $k_L$ values are decreasing in the following order: BaS $\textgreater$ BaO $\textgreater$ MgTe $\textgreater$ PbTe.}
\label{fig:PbTe}
\end{figure}

Out of 16 MX compounds, three of them (BaO, BaS and MgTe) found to have low $k_L$ ($\textless$ 6 Wm$^{-1}$K$^{-1}$) over the studied temperature range of 300-800 K. As illustrated in Figure \ref{fig:PbTe}, we compared phonon dispersion curves, phonon lifetimes, scattering rates and $k_L$ of BaO, BaS and MgTe compounds with PbTe. Softening of acoustic modes due to its high atomic mass (see Figure \ref{fig:PbTe}a), high scattering rates (see Figure \ref{fig:PbTe}b) and short phonon lifetimes (see Figure \ref{fig:PbTe}c) of PbTe responsible for its low $k_L$ behavior over BaO, BaS and MgTe compounds. The obtained $k_L$ values follow exactly the decreasing order of phonon lifetimes for these four compounds, which is given as follows: BaS $\textgreater$ BaO $\textgreater$ MgTe $\textgreater$ PbTe (see Figure \ref{fig:PbTe}c $\&$ d). Based on this trend and observed trends for other MX compounds (see Figure \ref{fig:LT}) clearly show that phonon lifetime ($\tau$) is a dominating factor to determine $k_L$ behavior in iso-structural compounds with the same crystal symmetry.

\begin{figure}
\centering
\subfigure[]{\includegraphics[width=3.0in,height=2.2in]{Figures/SBaO-PD.eps}} 
\subfigure[]{\includegraphics[width=3.2in,height=2.3in]{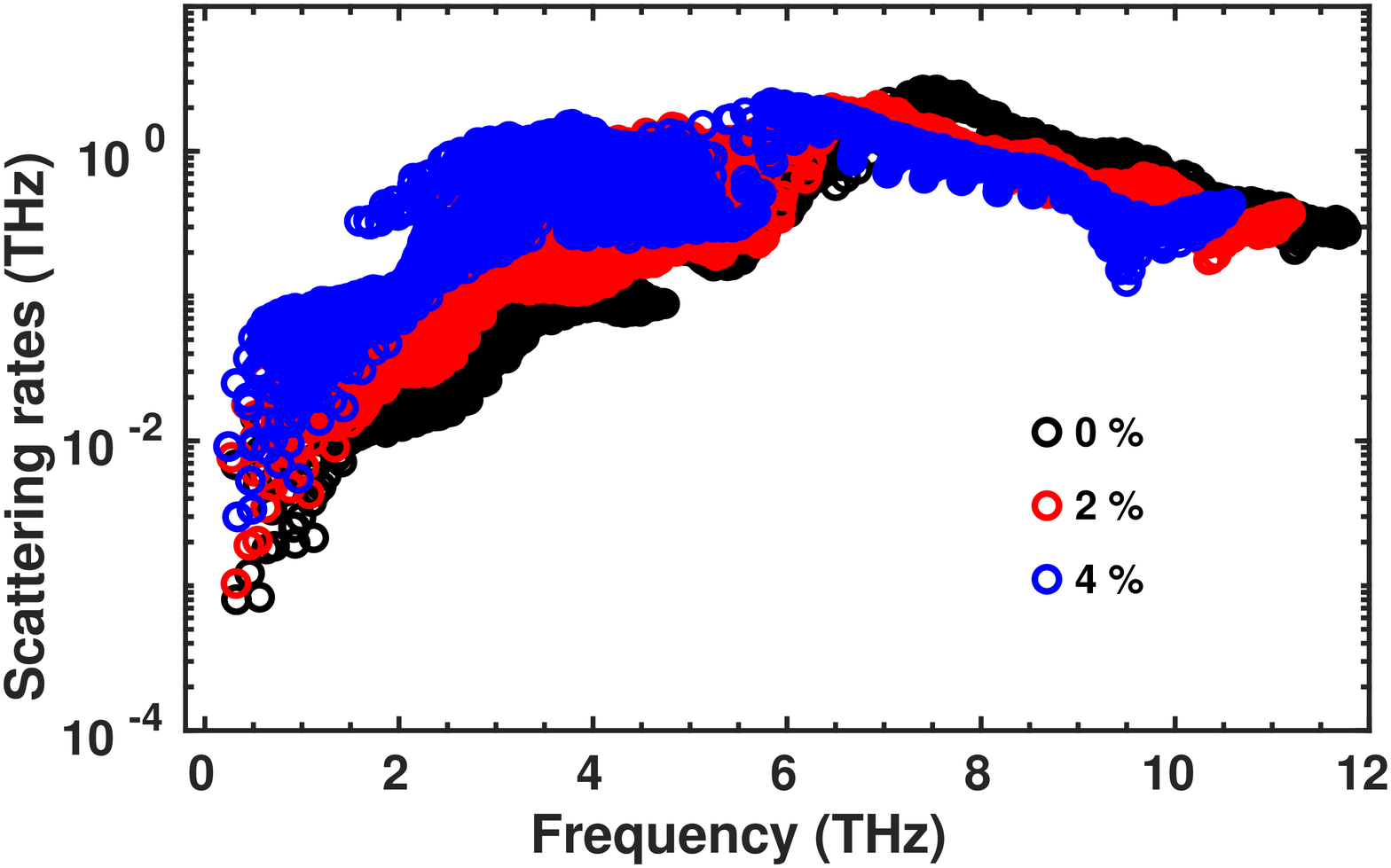}} \vspace{0.3in} \\
\subfigure[]{\includegraphics[width=3.2in,height=2.3in]{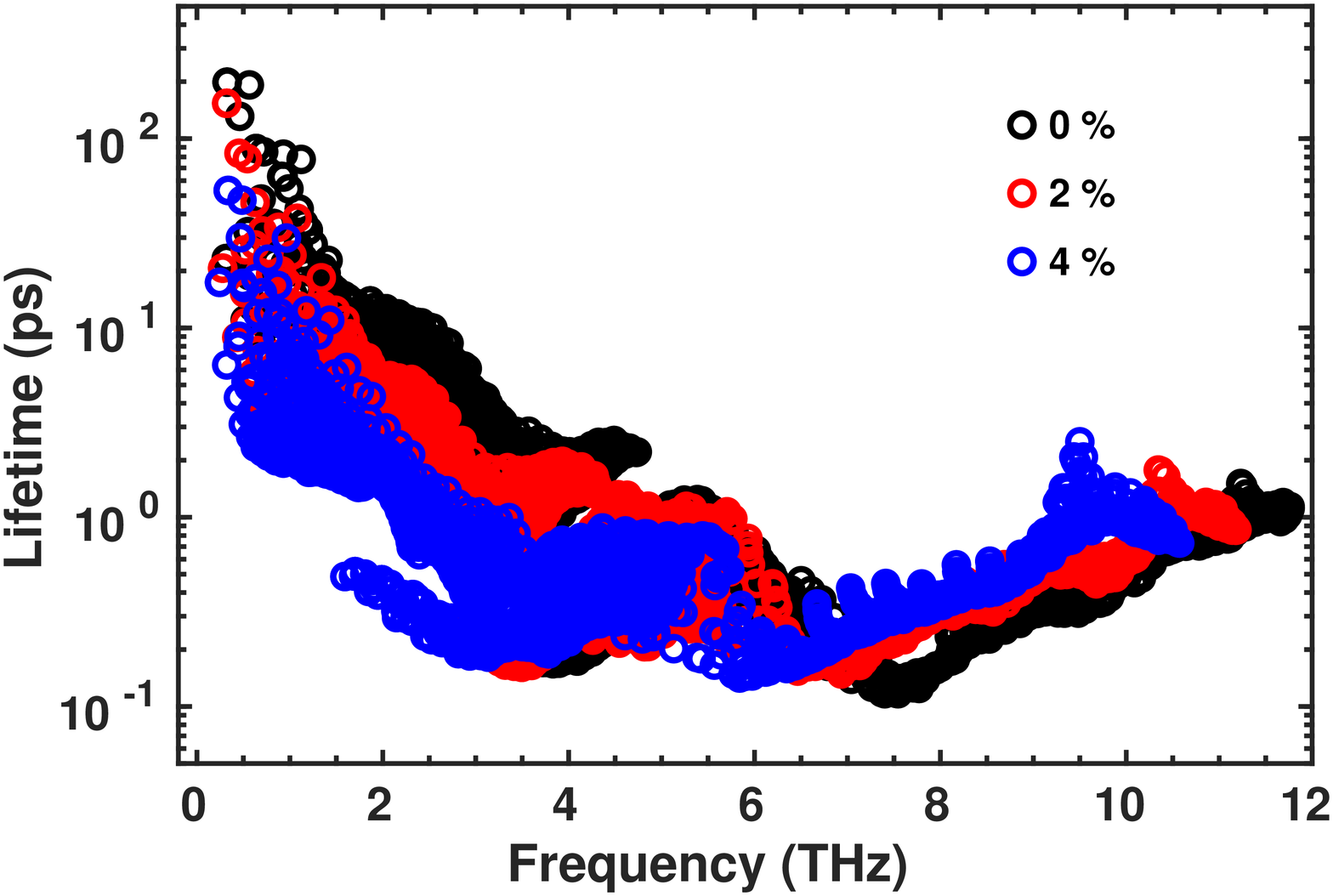}} 
\subfigure[]{\includegraphics[width=3.0in,height=2.3in]{Figures/SBaO-TC.eps}}
\caption{Calculated tensile strain dependent (a) phonon dispersion curves, (b) phonon scattering rates (c) phonon lifetime, and (d) lattice thermal conductivity ($k_L$) of BaO.}
\label{fig:SBaO}
\end{figure}

\begin{figure}
\centering
\subfigure[]{\includegraphics[width=3.0in,height=2.2in]{Figures/SBaS-PD.eps}} 
\subfigure[]{\includegraphics[width=3.2in,height=2.2in]{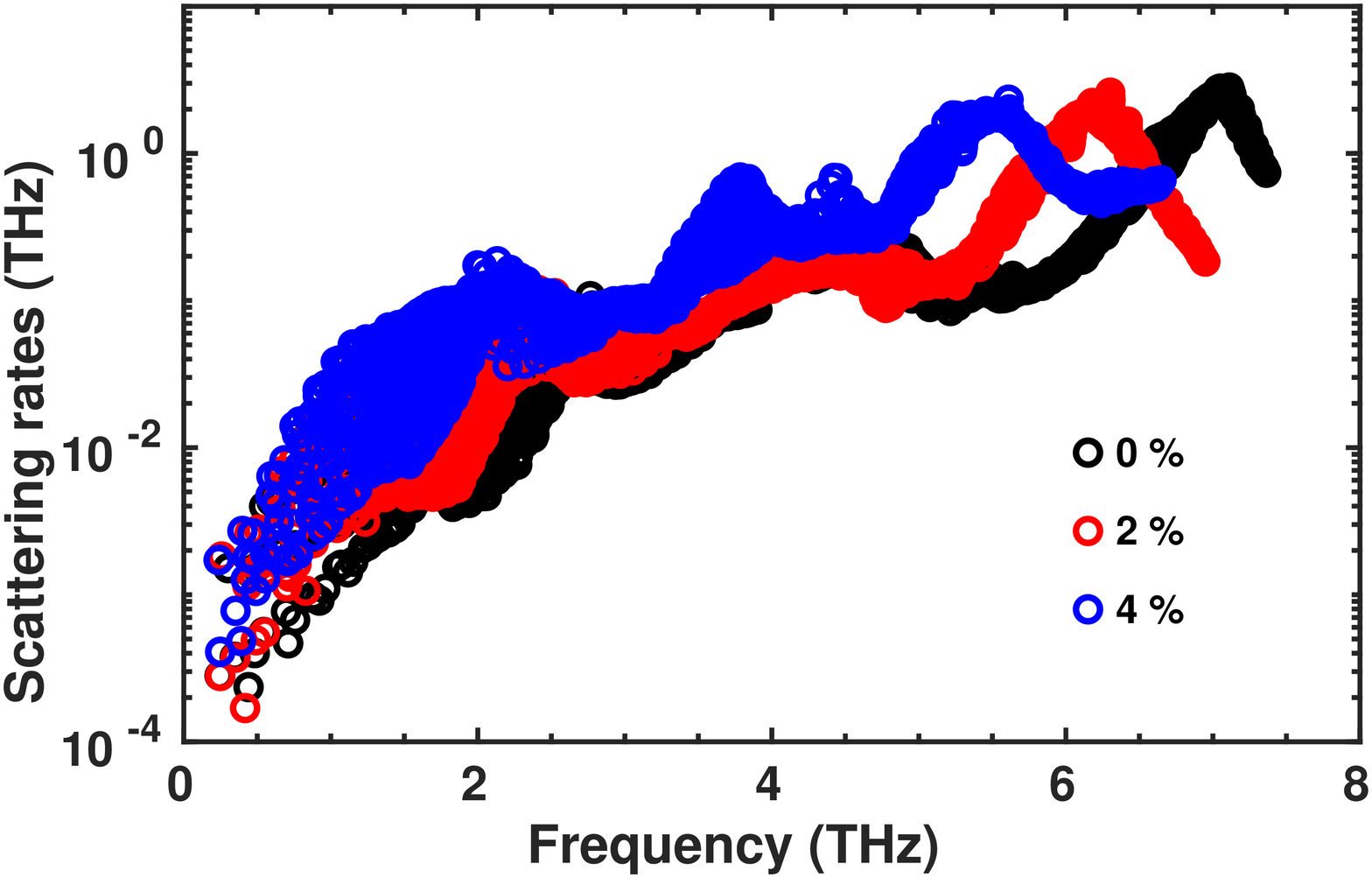}} \vspace{0.3in} \\
\subfigure[]{\includegraphics[width=3.2in,height=2.2in]{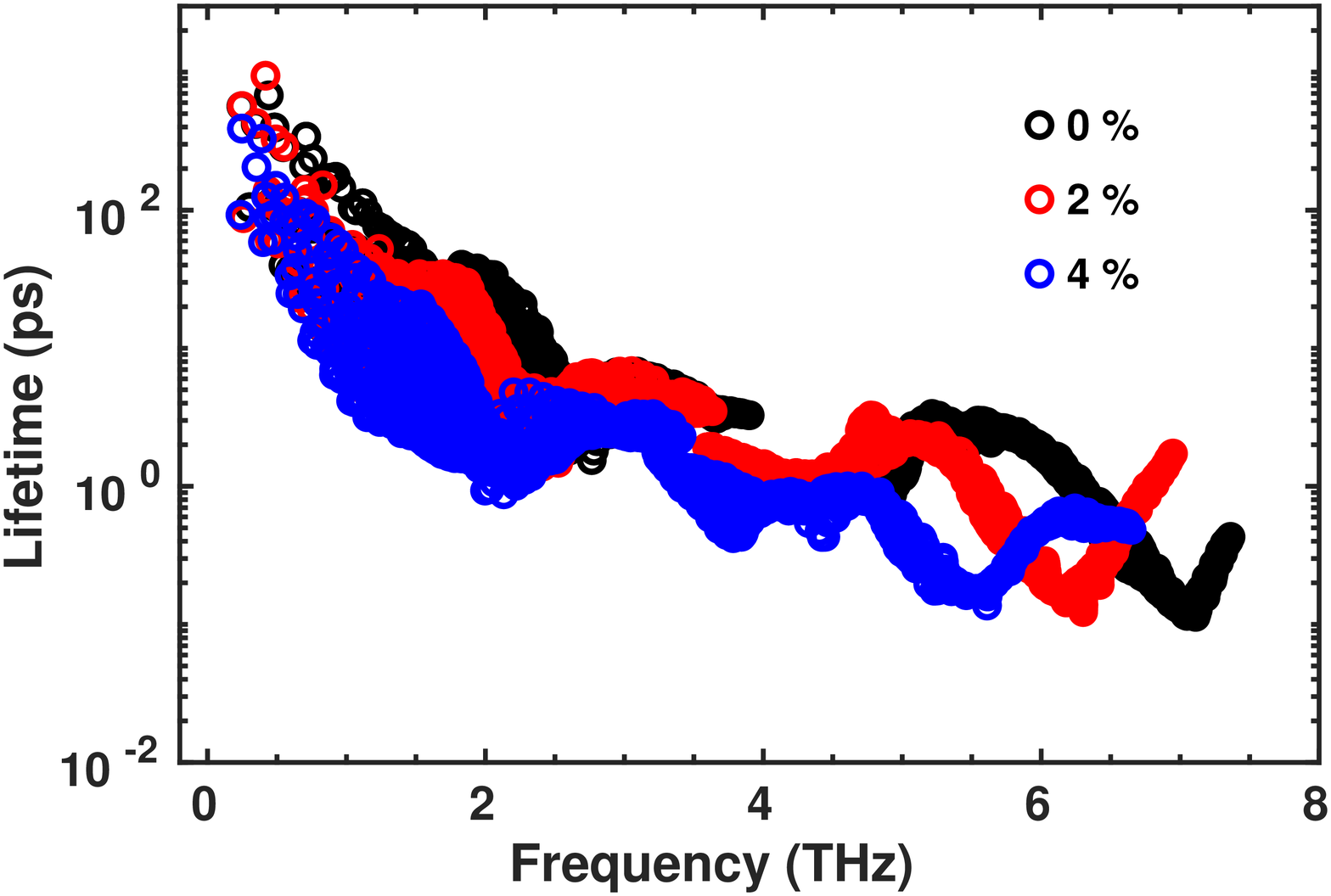}} 
\subfigure[]{\includegraphics[width=3.0in,height=2.2in]{Figures/SBaS-TC.eps}}
\caption{Calculated tensile strain dependent (a) phonon dispersion curves, (b) phonon scattering rates (c) phonon lifetime, and (d) lattice thermal conductivity ($k_L$) of BaS.}
\label{fig:SBaS}
\end{figure}

\begin{figure}
\centering
\subfigure[]{\includegraphics[width=3.0in,height=2.2in]{Figures/SMgTe-PD.eps}} 
\subfigure[]{\includegraphics[width=3.2in,height=2.2in]{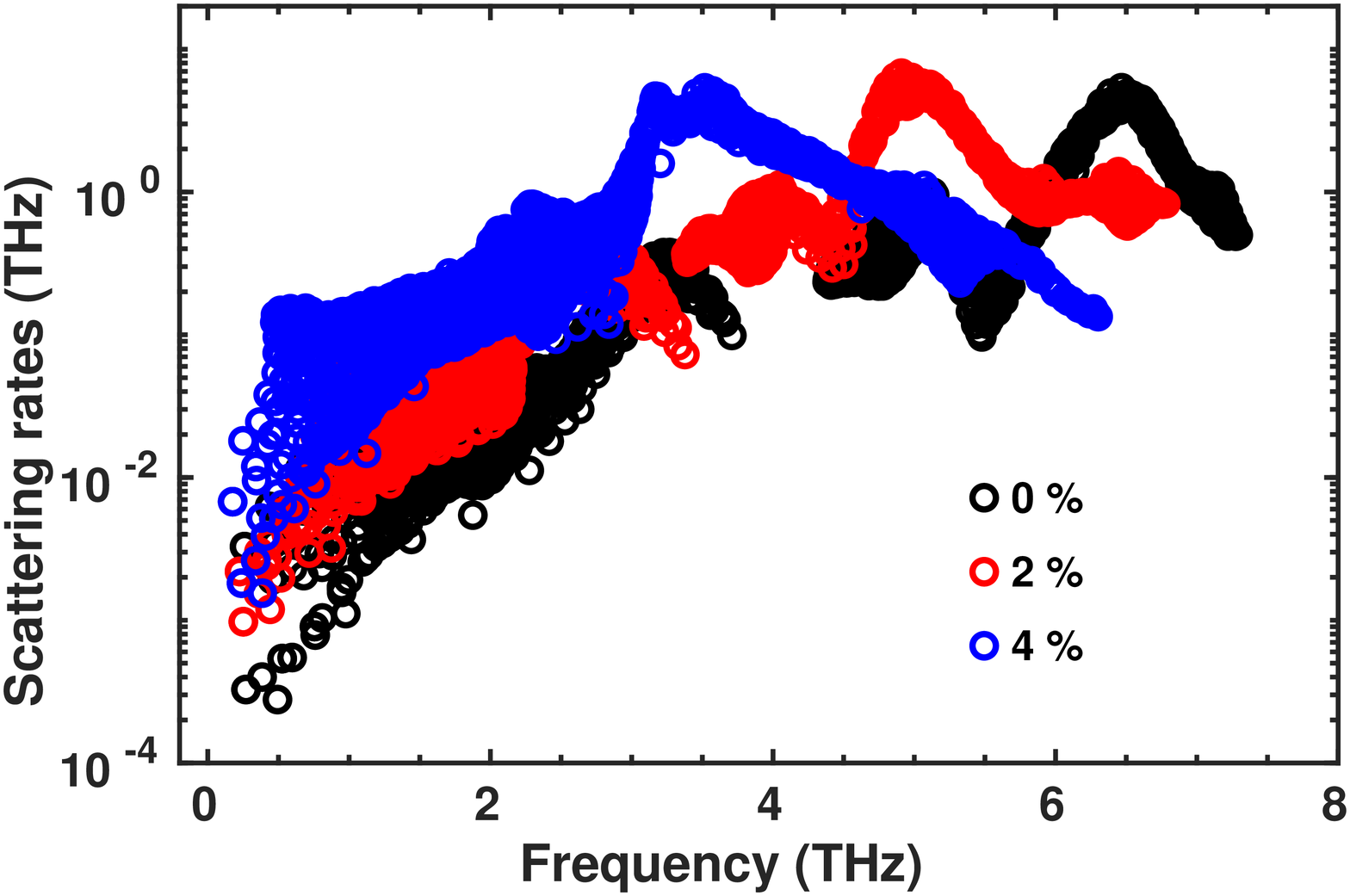}} \vspace{0.3in} \\
\subfigure[]{\includegraphics[width=3.2in,height=2.2in]{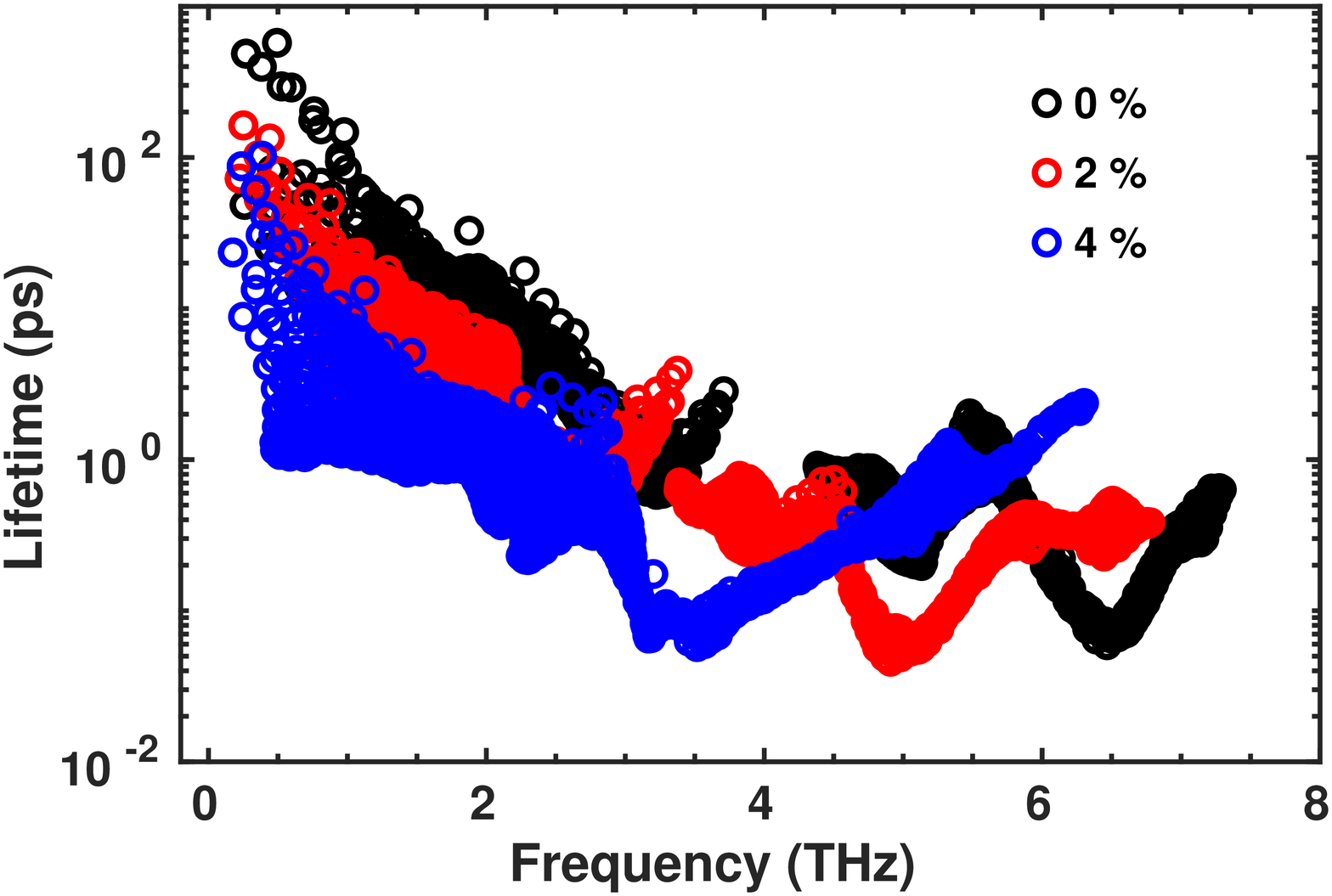}} 
\subfigure[]{\includegraphics[width=3.0in,height=2.2in]{Figures/SMgTe-TC.eps}}
\caption{Calculated tensile strain dependent (a) phonon dispersion curves, (b) phonon scattering rates (c) phonon lifetime, and (d) lattice thermal conductivity ($k_L$) of MgTe.}
\label{fig:SMgTe}
\end{figure}

We then considered these three BaO, BaS and MgTe compounds to investigate the effect of tensile strain on lattice dynamics and phonon transport. Further to lower $k_L$, we applied tensile strain, which is an effective strategy to achieve (ultra)low $k_L$ in materials. We have systematically increased the obtained equilibrium lattice constant up to 6$\%$ but we observed soft phonon modes with tensile strain $\geq$ 5$\%$ of the equilibrium lattice constant for BaO, therefore, we studied the effect of tensile strain up to 4$\%$ for these three compounds. As illustrated in Figures \ref{fig:SBaO}a, \ref{fig:SBaS}a $\&$ \ref{fig:SMgTe}a, with increasing strain, acoustic and TO  phonon modes are getting softened, which increases the coupling strength between acoustic and TO phonon modes. This eventually increases phonon-phonon scattering rates with increasing strain (see Figure \ref{fig:SBaO}b, \ref{fig:SBaS}b, \ref{fig:SMgTe}b), which causes a reduction in $k_L$ over the studied temperature range. The phonon lifetime decreases (see Figure \ref{fig:SBaO}c,  \ref{fig:SBaS}c, \ref{fig:SMgTe}c) significantly due to high scattering rates for both acoustic and low lying TO modes with increase in tensile strain which is responsible for further lowering of $k_L$ (see Figure \ref{fig:SBaO}d, \ref{fig:SBaS}d, \ref{fig:SMgTe}d). The obtained $k_L$ values for 4 $\%$ of tensile strain at 300 K are $\sim$ 2.06, $\sim$ 2.38, $\sim$ 1.05 Wm$^{-1}$K$^{-1}$ for BaO, BaS and MgTe, respectively. The (ultra)low $k_L$ of strained MgTe might be a better candidate for energy conversion applications. From the present and previous studies,\cite{Yang2020} one can expect a similar behavior for other MX compounds with application of tensile strain.  
\subsection{Elastic constants and mechanical properties}
To explore inter atomic bonding strength, lattice anharmonicity and mechanical stability of MX compounds, we have calculated second order elastic constants (C$_{ij}$). Since all the studied MX compounds crystallize in the cubic ($Fm\bar{3}m$) structure, they have three independent elastic constants such as longitudinal (C$_{11}$), transverse (C$_{12}$) and shear (C$_{44}$) due to symmetry constraints (C$_{11}$ = C$_{22}$ = C$_{33}$, C$_{12}$ = C$_{13}$ = C$_{23}$, C$_{44}$ = C$_{55}$ = C$_{66}$ and C$_{ij}$ = C$_{ji}$). The calculated second order elastic constants are given in Table S2 and are consistent with the available ultrasonic pulse echo\cite{CaSrO-Expt-elasticSON1972,MgO-Elastic0ExptANDERSON1966,BaO-elastic-Vetter1973} and Brillouin scattering measurements\cite{MgO-Expt-2-Sinogeikin1999} as well as with previous first principles calculations.\cite{MTeRen2017,MO-Elastic-CINTHIA201523,SrXAbdusSalam2019,CaX-RafikMaizi2019,MgSeNaClWu2015,Rajput2021,PbTe-a-elastic-Xue2021,Thermodynamic-1,CaTe-Guo2021} The obtained elastic constants satisfy the Born stability criteria\cite{Born-1,Born-2} indicating the mechanical stability of all these MX compounds.
\begin{equation}
C_{11} - C_{12} > 0, C_{11} > 0, C_{44} > 0, C_{11} +2C_{12} > 0
\end{equation}
We then computed bulk (B) and shear (G) moduli from the calculated elastic constants with Voigt-Reuss-Hill (VRH) approximation using equations 5 and 6, respectively. Later, B and G values are used to calculate Young's modulus (E) using equation 7. Since MgO has the highest E value, it is the stiffest material among the 16 MX compounds.
\begin{equation}
B =\frac{C_{11} +2C_{12}}{3}
\end{equation}
\begin{equation}
G =\frac{1}{2}\left [\frac{C_{11}-C_{12}+3C_{44}}{5} + \frac{5C_{44} (C_{11} - C_{12})} {4C_{44}+ 3(C_{11} - C_{12})}\right]
\end{equation}
\begin{equation}
E =\frac{9BG}{3B+G}
\end{equation}
The calculated C$_{ij}$'s, E, B, G moduli decrease from MO to MTe (M = Mg, Ca, Sr, Ba), which indicates the weak electrostatic/interatomic interactions in the lattice with increase in atomic size $i.e.,$ from Mg to Ba and O to Te. Therefore, the materials with higher atomic size can be easily deformed under mechanical stress thus results in low elastic moduli or soft lattice for systems with higher atomic mass. 
\begin{equation}
\sigma =\frac{3B-2G}{2(3B+G)}
\end{equation}
\begin{equation}
\gamma_\sigma = \frac{3}{2}\left(\frac{1+\sigma}{2-3\sigma}\right)
\end{equation}
The typical values of Poisson’s ratio ($\sigma$) are 0.1 and 0.25 for covalent and ionic materials, respectively.\cite{Sigma-Haines2001} The obtained $\sigma$ values span in the range of 0.18-0.28, which infer a strong ionic contribution in the interatomic bonding for these MX compounds (see Figure \ref{fig:ELF}). MgO (0.18) and BaO (0.28) have the smallest and largest $\sigma$ values, respectively among all the 16 MX compounds. The obtained $\sigma$ value 0.28 for BaO is close to the $\sigma$ of 0.25 (0.28\cite{Xiao2016}) for PbTe. Further, the $\sigma$ values are used to calculated the Gr\"uneisen parameter ($\gamma_\sigma$). 

\begin{figure}
\centering
\subfigure[]{\includegraphics[width=5.0in,height=3.5in]{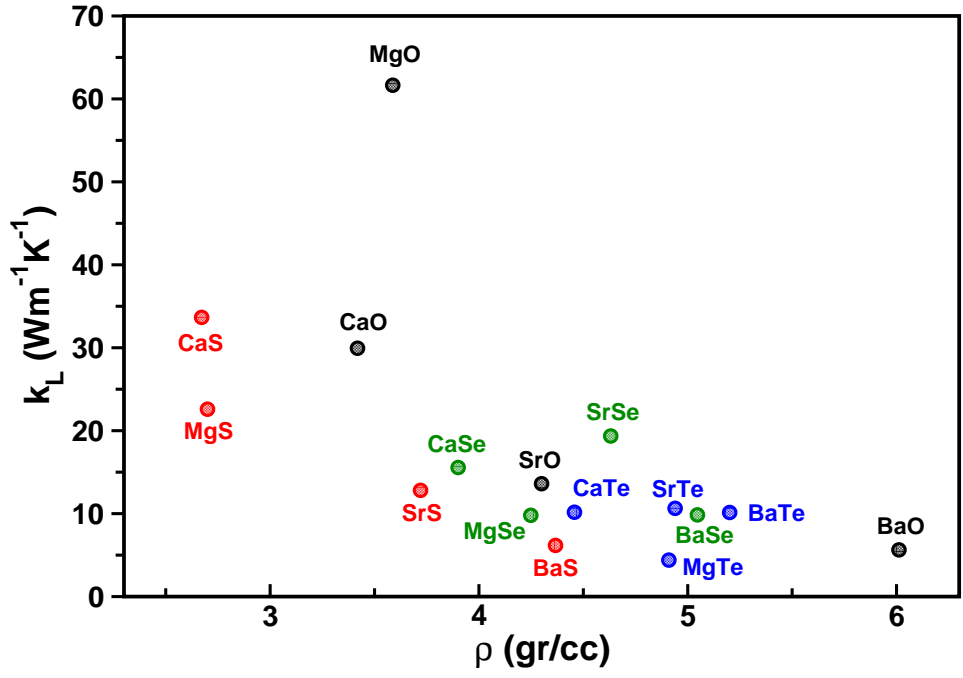}}  \vspace{0.5in} \\
\subfigure[]{\includegraphics[width=5.0in,height=3.5in]{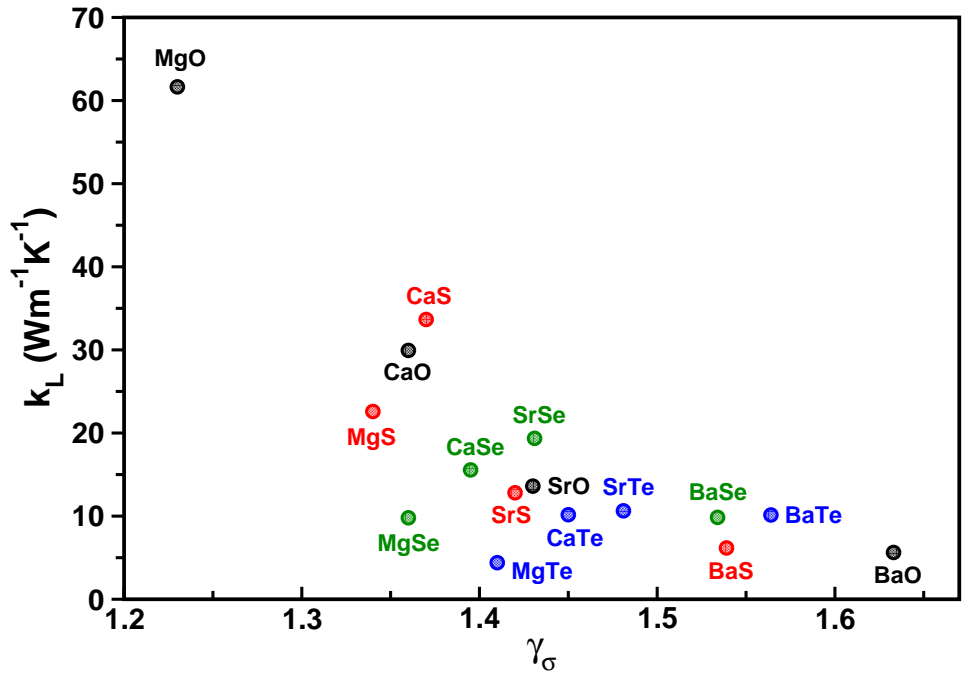}} 
\caption{Calculated $k_L$ as a function of (a) density ($\rho$) and (b) Gr\"uneisen parameter ($\gamma_\sigma$) for 16 MX compounds.}
\label{fig:gamma}
\end{figure}

\begin{figure}
\centering
\subfigure[]{\includegraphics[width=5.0in,height=3.5in]{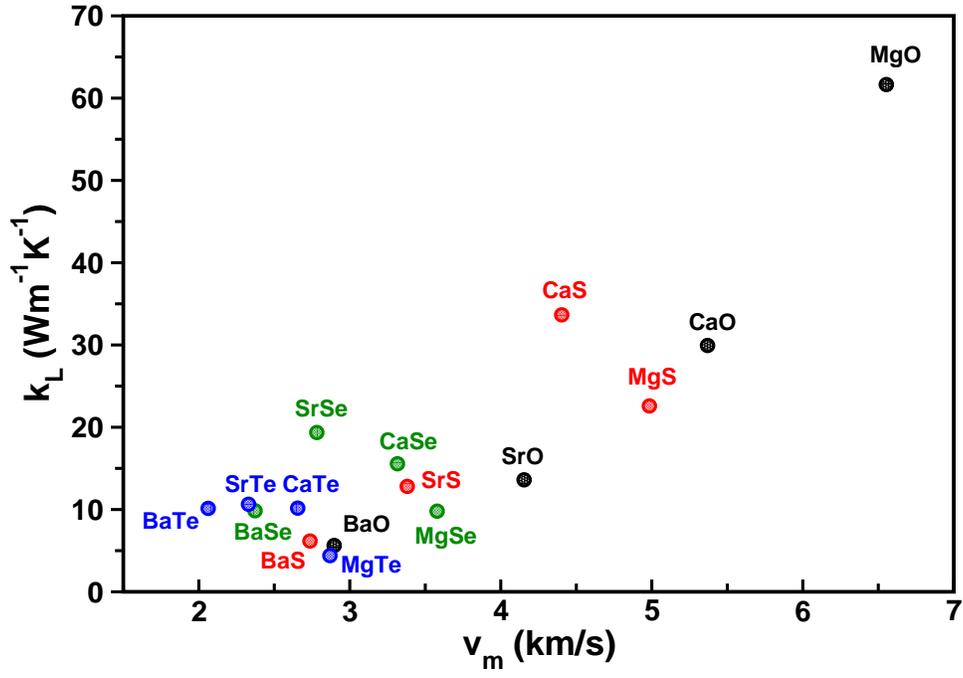}} \vspace{0.5in} \\
\subfigure[]{\includegraphics[width=5.0in,height=3.5in]{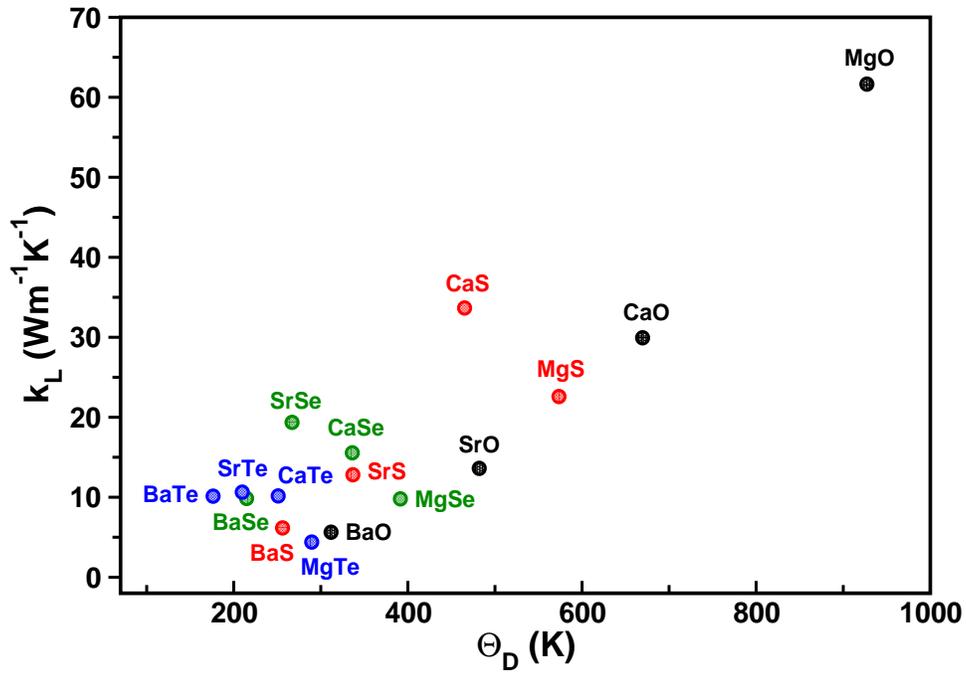}}
\caption{Calculated $k_L$ as a function of (a) average (v$_m$) sound velocity (b) Debye temperature ($\Theta_D$) for 16 MX compounds.}
\label{fig:Theta}
\end{figure}

Strength of lattice anharmonicity of a material is represented by Gr\"uneisen parameter ($\gamma_\omega$) which is usually obtained from phonons. However, computation of $\gamma_\omega$ involves a series of expensive phonon calculations and repeating them for 16 compounds is computationally demanding. To avoid this, we have used an efficient formula to compute $\gamma_\sigma$ using Poisson's ratio ($\sigma$)\cite{GP-Sanditov2011}. The obtained $\gamma_\sigma$ using equation 9 for materials with RS structure is in excellent agreement with $\gamma_\omega$.\cite{Xiao2016,Jia2017} Since the 16 materials under investigation crystallize in B1, we have used equation 9 to compute $\gamma_\sigma$ based on obtained $\sigma$ values with equation 8. Also, the obtained $\gamma_\sigma$ value 1.63 for BaO is closely comparable with $\gamma_\sigma$ 1.50 (1.65\cite{Xiao2016}) for PbTe. The deviation between this work and previous\cite{Xiao2016} studies might be due to distinct lattice constant used for computation of elastic constants, which are highly sensitive to the lattice constant used in the calculations. As shown in Figure \ref{fig:gamma}, BaO has the highest $\gamma_\sigma$, which indicates relatively high anharmonicity of BaO over other MX compounds which in turn leads to low $k_L$. We then calculated sound velocities (v$_l$, v$_t$, v$_m$) and Debye temperature ($\Theta_D$) using the following relationships:
\begin{equation}
v_l = \sqrt \frac{B+\frac{4G}{3}}{\rho}
\end{equation}
\begin{equation}
v_t = {\sqrt \frac{G}{\rho}}
\end{equation}
\begin{equation}
v_m = \left[\frac{1}{3} \left(\frac{1}{v_l^3}\right) +  \left(\frac{2}{v_t^3}\right)\right]^{\frac{-1}{3}}
\end{equation}
\begin{equation}
\Theta_D =  \frac{h}{k_B}\left[\left(\frac{3N}{4\pi V}\right)\right]^{\frac{1}{3}}v_m
\end{equation}
Here, $\rho$, h, k$_B$, N and V are the crystal density, Planck constant, Boltzmann constant, and number of atoms and volume of unit cell. \\
The calculated v$_l$, v$_t$, v$_m$ and $\Theta_D$ are decreasing from MO to MTe (M = Mg, Ca, Sr, Ba). Figures $\ref{fig:Theta}$ and S7 show the variation of $k_L$ as a function of Debye temparature ($\Theta_D$), average (v$_m$), and longitudinal (v$_l$), transverse (v$_t$) sound velocities for the 16 MX compounds. The same trend is observed for all these four properties v$_l$, v$_t$, v$_m$ and $\Theta_D$. According to the Slack theory, low $\theta_D$ are indicative of low $k_L$ in materials. In fact, the presence of LLO phonons results in a softening of the acoustic phonon modes, which yields low group velocities and frequencies for acoustic phonons, thus results in low $\theta_D$. However, BaO, BaS and MgTe compounds have low $k_L$ despite their higher v$_l$, v$_t$, v$_m$ and $\Theta_D$ over other MTe (M = Ca, Sr, Ba), BaSe, SrSe compounds. Moreover, the calculated phonon group velocities for BaX compounds (see Figure S8) also follow their atomic mass trend consistent with sound velocities. This results strongly suggest that phonon lifetime is the dominant factor over group velocity which is mainly responsible for the observed anomalous trends in MX compounds along with LLO and low frequency acoustic modes.

\section{Conclusions}
In summary, we have systematically investigated lattice dynamics, phonon transport and mechanical properties of 16 binary systems with rocksalt-type structure and compared their properties with an efficient thermometric material, PbTe. We predicted anomalous trends for $k_L$ in CaX (CaS $\textgreater$ CaO $\textgreater$ CaSe $\textgreater$ CaTe), SrX (SrSe $\textgreater$ SrO $\textgreater$ SrS $\textgreater$ SrTe ) and BaX (BaTe $\textgreater$ BaSe $\textgreater$ BaS $\textgreater$ BaO) series of compounds. In particular, we observed an opposite trend for $k_L$ in BaX series  which is in contrast to the expected trend from their atomic masses. However, the above trends are slightly altered for SrX and BaX compounds, when $k_L$ is computed at the experimental lattice constant. The possible reasons for such low $k_L$ behavior in BaO, BaS and MgTe compounds might be due to the following: 1) softening of acoustic (TA) phonon modes despite of low atomic mass 2) low lying TO phonon modes fall deep into acoustic mode region which enhances phonon scattering phase space, 3) short phonon lifetimes and high scattering rates, 4) high density ($\rho$) and large Gr\"uneisen parameter ($\gamma_\sigma$). Application of tensile strain further reduces $k_L$ in BaO, BaS and MgTe through phonon softening which increases scattering rates thereby lowering phonon lifetimes. Overall, the present study provides insights to achieve (ultra)low $k_L$ materials through phonon engineering in simple crystal systems, which is essential for the development of sustainable energy conversion devices for future energy applications.

\section{Acknowledgments}
SCRR and NYK contributed equally to the manuscript. SCRR would like thank RGUKT Basar for providing computational facilities. NYK would like to thank Prof. Olle Hellman, Link\"oping University, Sweden for his valuable suggestions. NYK also would like to thank Science and Engineering Research Board and Indo-US Scientific Technology Forum for providing financial support through SERB Indo-US postdoctoral fellowship and Stony Brook Research Computing and Cyberinfrastructure, and the Institute for Advanced Computational Science at Stony Brook University for access to the high-performance SeaWulf computing system, which was made possible by a $\$$1.4M National Science Foundation grant ($\#$1531492). 

\clearpage
\bibliography{Refs.bib}

\clearpage

\begin{table}[tbp]
\caption{Calculated equilibrium lattice constant (a, in \AA) of MX compounds at ambient pressure and are compared with available X-ray diffraction data and other first principles calculations.}
\label{table1}
\begin{tabular}{ccccc} \hline
Compound   &   This work    &  Expt.     &  Others        \\ \hline
MgO        &     4.210 &        4.21$^a$     &  4.165$^b$,4.2130$^c$     &   \\    
MgS        &     5.176 &  5.2033$^d$  & 5.19$^e$ ,5.234$^f$ ,5.2018$^g$ \\
MgSe       &     5.445 & 5.463$^h$  & 5.46$^e$ ,5.512$^f$ ,5.4823$^g$\\ 
MgTe       &     5.901 & -  & 5.98$^e$ ,5.980$^f$ ,5.9585$^g$\\
CaO        &  4.776   & 4.81$^i$  &4.72$^b$ ,4.828$^j$ \\    
CaS        &  5.639   & 5.689$^k$ & 5.67$^l$ ,5.712$^j$,5.637$^m$    \\
CaSe       &  5.874    & 5.916$^k$  & 5.91$^l$ ,5.965$^j$,5.868$^m$   \\
CaTe       &  6.298   & 6.348$^k$ ,6.356$^n$ & 6.33$^l$ ,6.399$^j$ ,6.4430$^o$ \\ 
SrO        &  5.134 & 5.16$^p$ & 5.073$^b$ ,5.184$^q$  \\    
SrS        &  5.978 & 6.024$^r$   & 6.076$^s$ ,6.05 $^q$\\
SrSe       &  6.205 & 6.236$^k$    & 6.323$^s$ ,6.3$^q$  \\ 
SrTe       &  6.614 & 6.66$^t$ ,6.659$^n$  & 6.76$^s$ ,6.718$^q$ ,6.66$^u$  ,6.683$^u$   \\ 
BaO        &  5.533 &  5.520$^v$  &  5.562$^w$, 5.454$^w$ ,5.614$^x$ \\    
BaS        &  6.363 &  6.387$^y$ & 6.44$^z$, 6.38998$^{aa}$ ,6.407$^w$  ,6.455$^x$  \\
BaSe       &  6.578 &  6.59$^{ab}$ & 6.59298$^{aa}$ ,6.640$^w$ ,6.699$^x$ ,6.668$^{ac}$ \\ 
BaTe       &  6.968 &  6.99$^{ab}$   & 7.00598$^a_a$ ,6.989$^w$ ,7.080$^x$ ,7.075$^{ac}$ \\ 
PbTe     &   6.441      &  6.46$^{ad}$ & 6.45$^{ae}$, 6.35$^{af}$,6.576$^{ag}$\\ \hline
\end{tabular}
\newline
$^a$Ref.\cite{MgO-Expt} $^b$Ref.\cite{MO-Theory} $^c$Ref.\cite{MgOaFu2017} $^d$Ref.\cite{MgS-Expt} $^e$Ref. \cite{MgX-Theory-Khan2012} $^f$Ref.\cite{MgX-Theory-Tairi2017} $^g$Ref.\cite{MgX-Mir2016}$^h$Ref.\cite{MgSe-Expt} $^i$Ref.\cite{CaO-Expt} $^j$Ref.\cite{Rajput2019} $^k$Ref. \cite{CaX-Expt} $^l$Ref.\cite{CaS-Se-Te-1} $^m$Ref.\cite{Mg-CaX-Debnath2018} $^n$Ref.\cite{CaTe-SrTe-EXPT} $^o$Ref.\cite{CaTe-Theory-CihanKrk2019} $^p$Ref.\cite{SrO-Expt} $^q$Ref.\cite{Rajput2020} $^r$Ref. \cite{SrS-Expt} $^s$Ref.\cite{SrX-a-Theory} $^t$Ref.\cite{SrTe-Exp} $^u$Ref.\cite{SrTe-lattice-Saoud2016} $^v$ Ref.\cite{BaO-Expt} $^w$Ref. \cite{BaX-Theory-Lin2005} $^x$Ref.\cite{Rajput2021} $^y$Ref.\cite{BaS-Expt} $^z$Ref.\cite{BaS-2017} $^{aa}$Ref.\cite{BaX-Theory-Kholiya2011}  $^{ab}$Ref.\cite{BaSe-Te-Expt}         $^{ac}$Ref.\cite{BaSe-Te-Drablia2017} $^{ad}$Ref. \cite{PbTe-Expt-lattice-Bouad2003} $^{ae}$Ref.\cite{PbTe-Zhang-Theory-a-2021} $^{af}$Ref.\cite{elastic-PbTe-Yang2012} $^{ag}$Ref.\cite{PbTe-a-elastic-Xue2021}
\end{table}

\begin{table}[tbp]
\caption{Calculated  lattice thermal conductivity ($k_L$, in Wm$^{-1}$K$^{-1}$) at 300 K for 16 MX and PbTe compounds.}
\label{table2}
\begin{tabular}{ccccccccc} \hline
          &     This work & &        & Others  \\  \hline
Compound  &  $k_L$ & $k_L^E$ & Expt.  & $k^{HA}_{3ph}$ & $k^{SCPH}_{3ph}$ &  $k^{SCPH}_{3,4ph}$ & Others \\ \hline
MgO 	  &	61.65  & 61.65  & 52$^a$ & 52.1$^b$       &  58.7$^b$   &  50.1$^b$ &  -  \\ 
MgS       &	22.59  & 18.74  &  -     &   -            &    -        &    -      &  - \\ 
MgSe      &	9.8    &  7.55  &  -     &   -            &    -        &    -      &  -   \\ 
MgTe      &	4.45   &  4.45  &  -     &   -            &    -        &    -     & 3.5$^c$, 3$^d$  \\ 
CaO       &	29.94  &  24.77 & 30$^a$ &   21.3$^b$     & 25.1$^b$    & 22.2$^b$ & - \\ 
CaS	      &	33.66  & 28.39  &  -     &     -          &   -         &    -      &   -  \\ 
CaSe      &	15.56 & 13.21   &  -     &     -          &   -         &    -      &   -  \\ 
CaTe      &	10.17 & 8.33    &   -    &     -          &   -         &    -      & 3.1$^c$,8.5$^d$\\ 
SrO       &	13.61 & 11.65   & 10$^a$ &  9.0$^b$     & 11.0$^b$      & 9.9$^b$   & -   \\ 
SrS	      &	12.81 & 11.60   &   -    &     -          &    -        &    -      &   -  \\ 
SrSe      &	19.36 & 15.8    &   -    &     -          &    -        &    -      &   -  \\  
SrTe      &	10.64 & 9.66    &   -    &      -         &    -        &     -     & 2.5$^c$,10.5$^d$ \\
BaO  	  & 5.63 & 6.76    &  3$^a$ &   2.8$^b$      & 4.4$^b$     & 3.3$^b$   &  - \\ 
BaS	  	  & 6.17 & 6.28   &   -    &     -          &    -        &    -      &   -  \\  
BaSe      &	9.85 & 10.06  &   -    &     -          &    -        &    -      &   -  \\   
BaTe      & 10.14 & 9.84  &   -    &     -          &    -        &    -     & 1.9$^c$,10.2$^d$ \\ 
PbTe      & 3.93 & 3.21   & 3.1$^e$, 3.0$^e$ & - & 3.3$^d$ & 2.6$^f$ & -    \\ 
          &  -    & -   &  2.38$^e$, 2.6$^e$ &   - &     -  &    -    & - \\ \hline
\end{tabular}
\newline
$^a$Ref.\cite{Kl-Expt-Ref-1999} $^b$Ref.\cite{Kl-Ref-Xia2020} $^c$ Ref.\cite{MTeRen2017} $^d$Ref.\cite{Xia2018} $^e$Ref.\cite{PbTe-Kl-Expt-Bagieva2012} $^f$Ref.\cite{PbTe-Different-Xia2018} \\
$k_L$: calculated at the PBEsol equilibrium lattice constant  \\
$k_L^E$: calculated at the experimental lattice constant  \\
\end{table}

\end{document}


\subsection*{Lattice thermal conductivity of MX compounds}
\begin{figure}
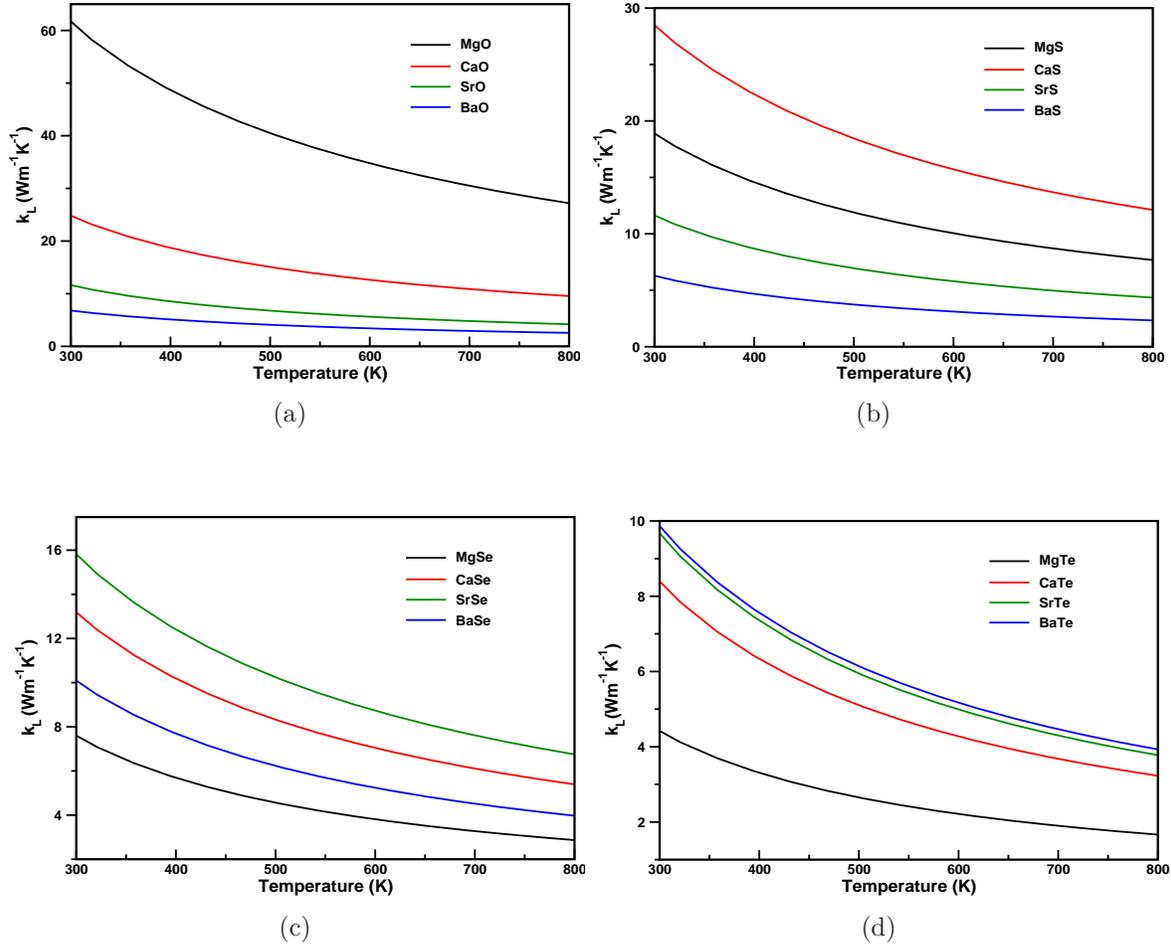

\subfigure[]{\includegraphics[width=3.0in,height=2.0in]{Figures/MO-TC.eps}} 
\subfigure[]{\includegraphics[width=3.0in,height=2.0in]{Figures/MS-TC.eps}} \vspace{0.3in} \\
\subfigure[]{\includegraphics[width=3.0in,height=2.0in]{Figures/MSe-TC.eps}} 
\subfigure[]{\includegraphics[width=3.0in,height=2.0in]{Figures/MTe-TC.eps}}
\caption{Calculated lattice thermal conductivity of (a) MO, (b) MS, (c) MSe and (d) MTe compounds as a function of temperature at the experimental lattice constant; where M = Mg, Ca, Sr and Ba.}
\label{fig:TC-M}
\end{figure}

\begin{figure}
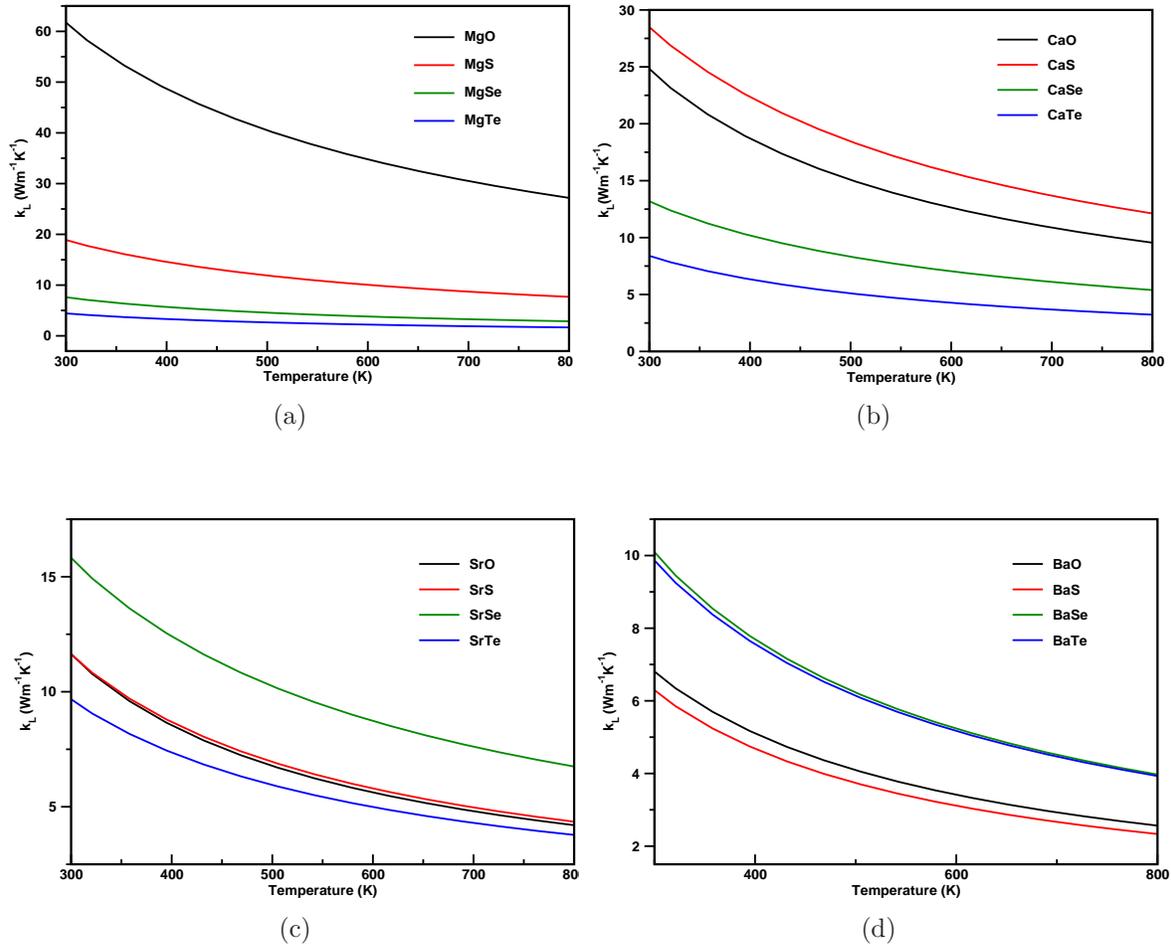

\centering
\subfigure[]{\includegraphics[width=3.0in,height=2.0in]{Figures/EMgX-TC.eps}} 
\subfigure[]{\includegraphics[width=3.0in,height=2.0in]{Figures/ECaX-TC.eps}} \vspace{0.3in} \\
\subfigure[]{\includegraphics[width=3.0in,height=2.0in]{Figures/ESrX-TC.eps}} 
\subfigure[]{\includegraphics[width=3.0in,height=2.0in]{Figures/EBaX-TC.eps}}
\caption{Calculated lattice thermal conductivity of (a) MgX, (b) CaX, (c) SrX and (d) BaX compounds as a function of temperature at the experimental lattice constant; where X = O, S, Se and Te.}
\label{fig:TC-X}
\end{figure}

We have also calculated lattice thermal conductivity ($k_L$) for MX compounds at the experimental lattice constant (see Table 2) by varying metal cation with fixed chalcogen atoms (see Figure \ref{fig:TC-M}) and vice versa (see Figure \ref{fig:TC-X}). As shown in Figure \ref{fig:TC-M}, the $k_L$ values are decreasing with increasing atomic mass $i.e.,$ from MgO $\textgreater$ CaO $\textgreater$ SrO $\textgreater$ BaO in MO compounds. While anomalous trends are observed between MgS and CaS with the following trend for MS (CaS $\textgreater$ MgS $\textgreater$ SrS $\textgreater$ BaS) compounds, for MSe (SrSe $\textgreater$ CaSe $\textgreater$ BaSe $\textgreater$ MgSe) and MTe (BaTe $\textgreater$ SrTe $\textgreater$ CaTe $\textgreater$ MgTe) compounds. Especially, MTe series shows an opposite trend for $k_L$ in contrary to the expected trend from their atomic mass. 

\clearpage

\begin{figure}
\includegraphics[width=5.0in,height=3.4in]{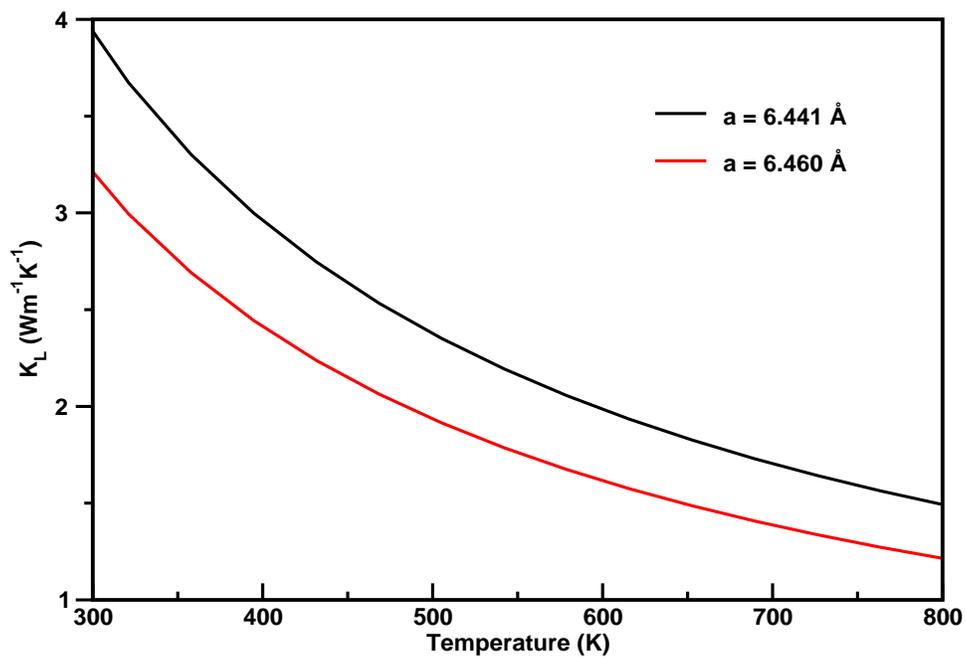}
\caption{Calculated lattice thermal conductivity of PbTe at equilibrium lattice constant (a = 6.441 $\AA$) using PBEsol and experimental lattice constant (a = 6.460 $\AA$). This clearly shows sensitivity of $k_L$ against lattice constant for PbTe.}
\label{fig:Theta}
\end{figure}

\begin{figure}
\centering
\subfigure[]{\includegraphics[width=3.0in,height=2.0in]{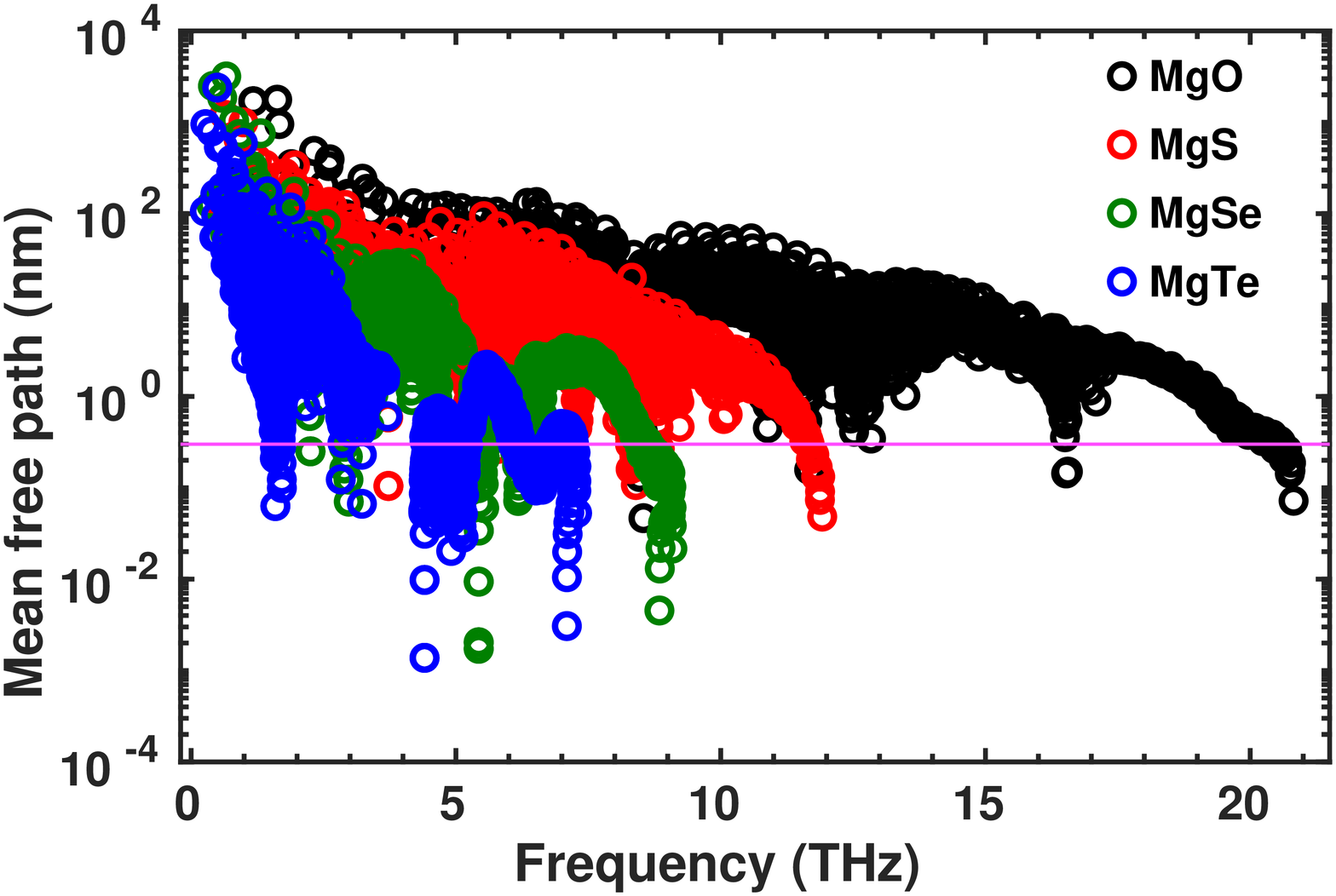}} 
\subfigure[]{\includegraphics[width=3.0in,height=2.0in]{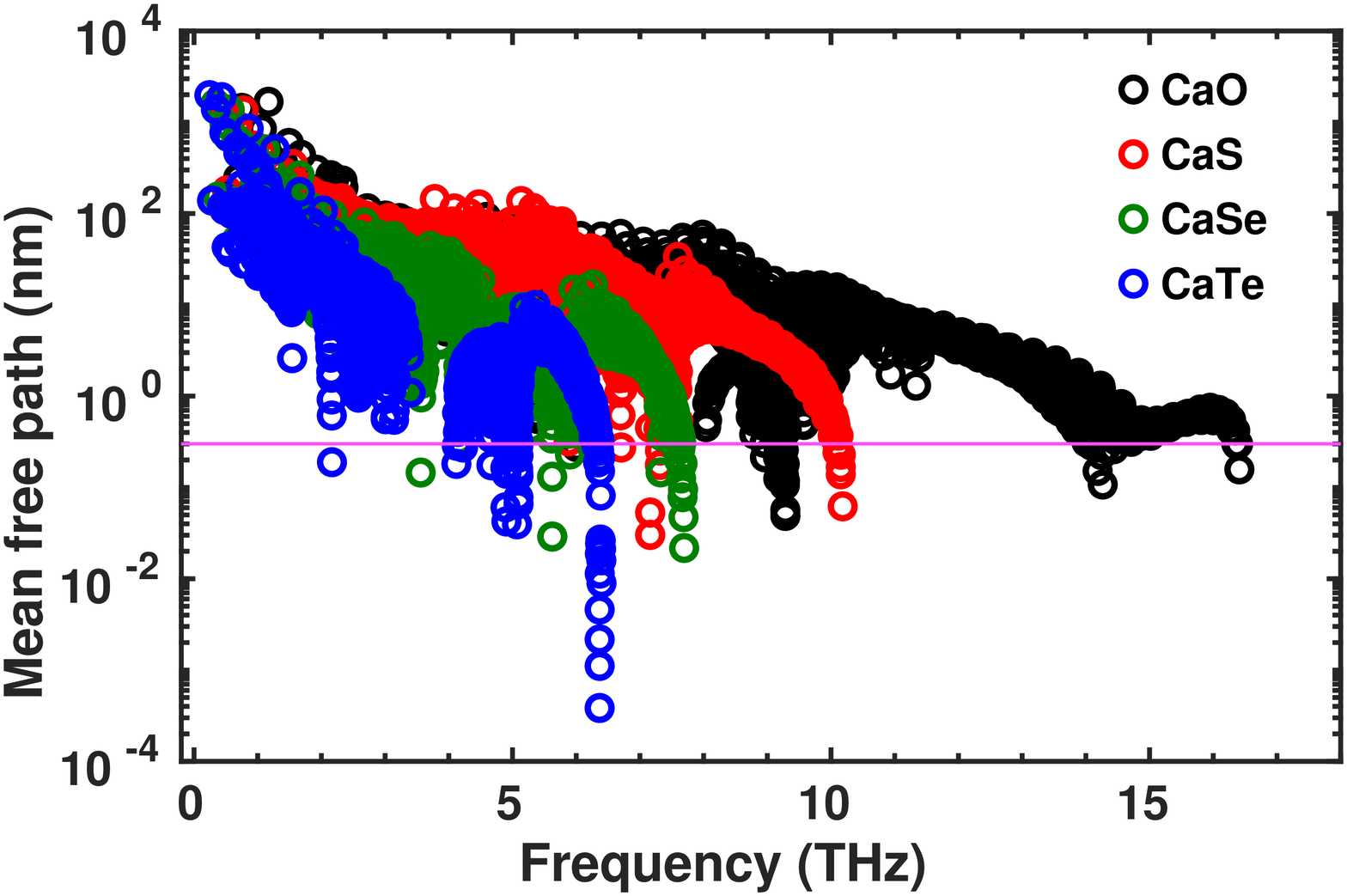}} \vspace{0.3in} \\
\subfigure[]{\includegraphics[width=3.0in,height=2.0in]{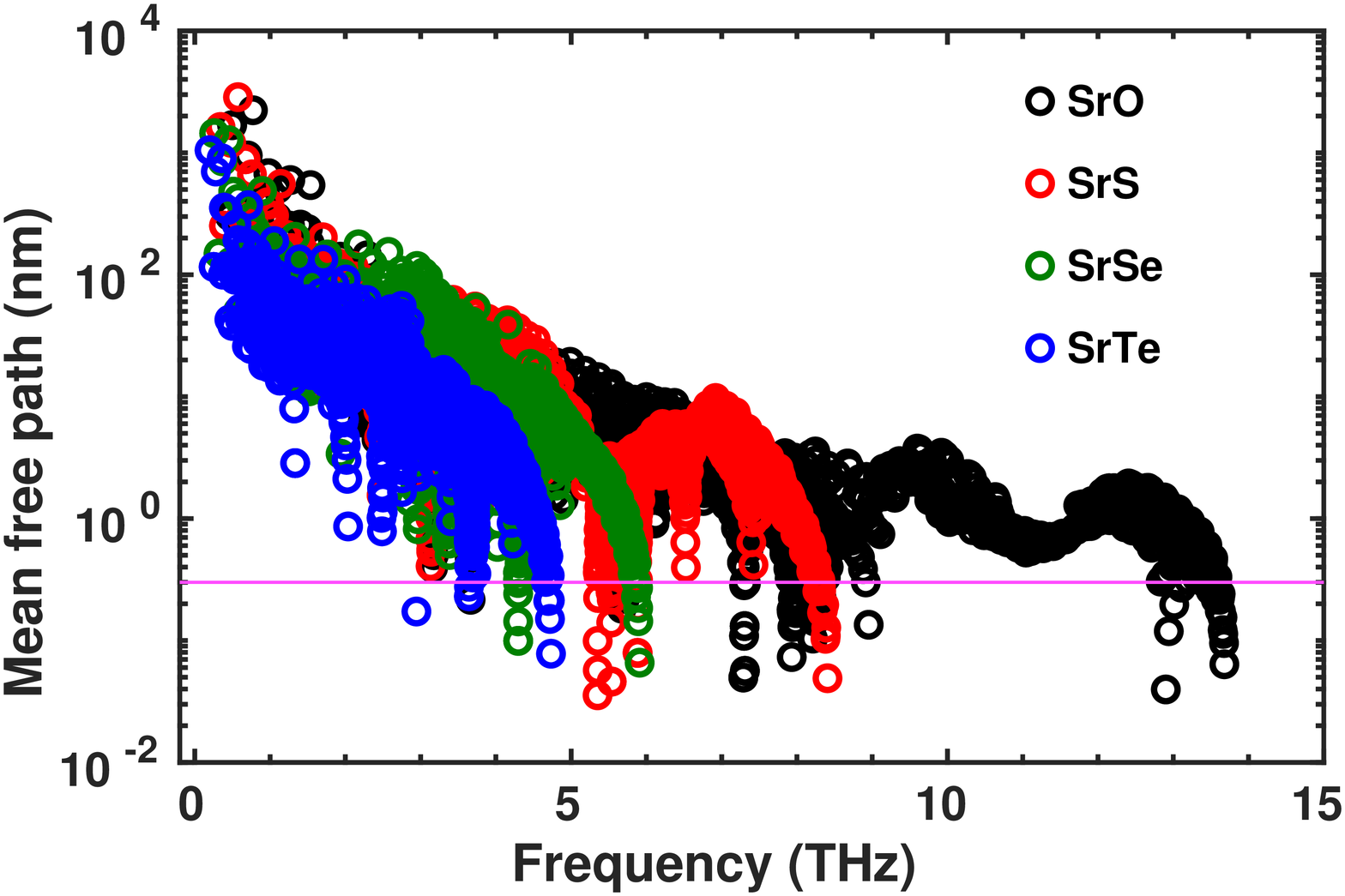}} 
\subfigure[]{\includegraphics[width=3.0in,height=2.0in]{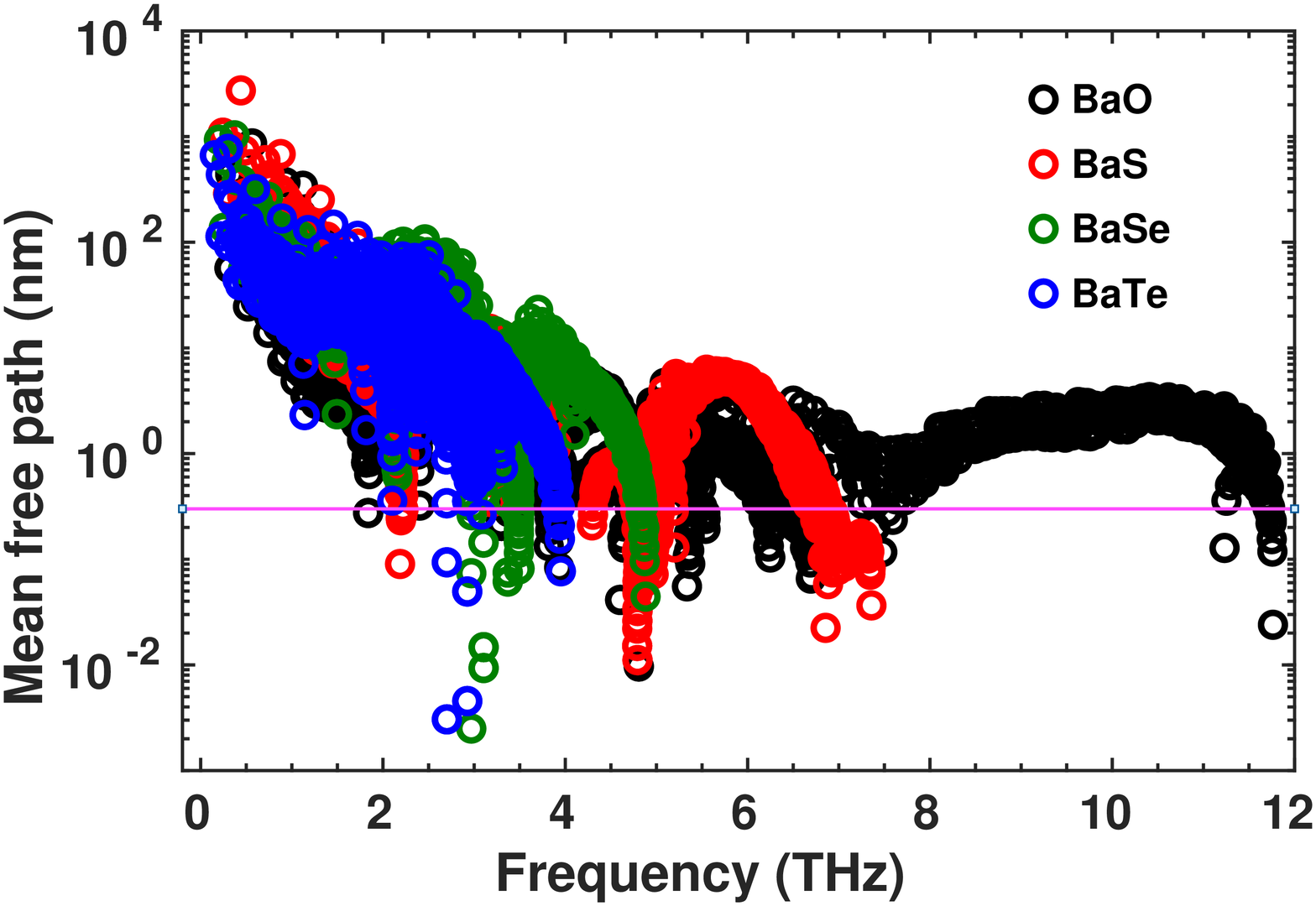}}
\caption{Calculated mean free paths of (a) MgX, (b) CaX, (c) SrX and (d) BaX compounds as a function of frequency; where X = O, S, Se and Te. The horizontal solid line (magenta) represents Ioffe-Regel limit for mean free paths.}
\label{fig:MFP}
\end{figure}

\begin{figure}
\centering
\subfigure[]{\includegraphics[width=3.0in,height=2.0in]{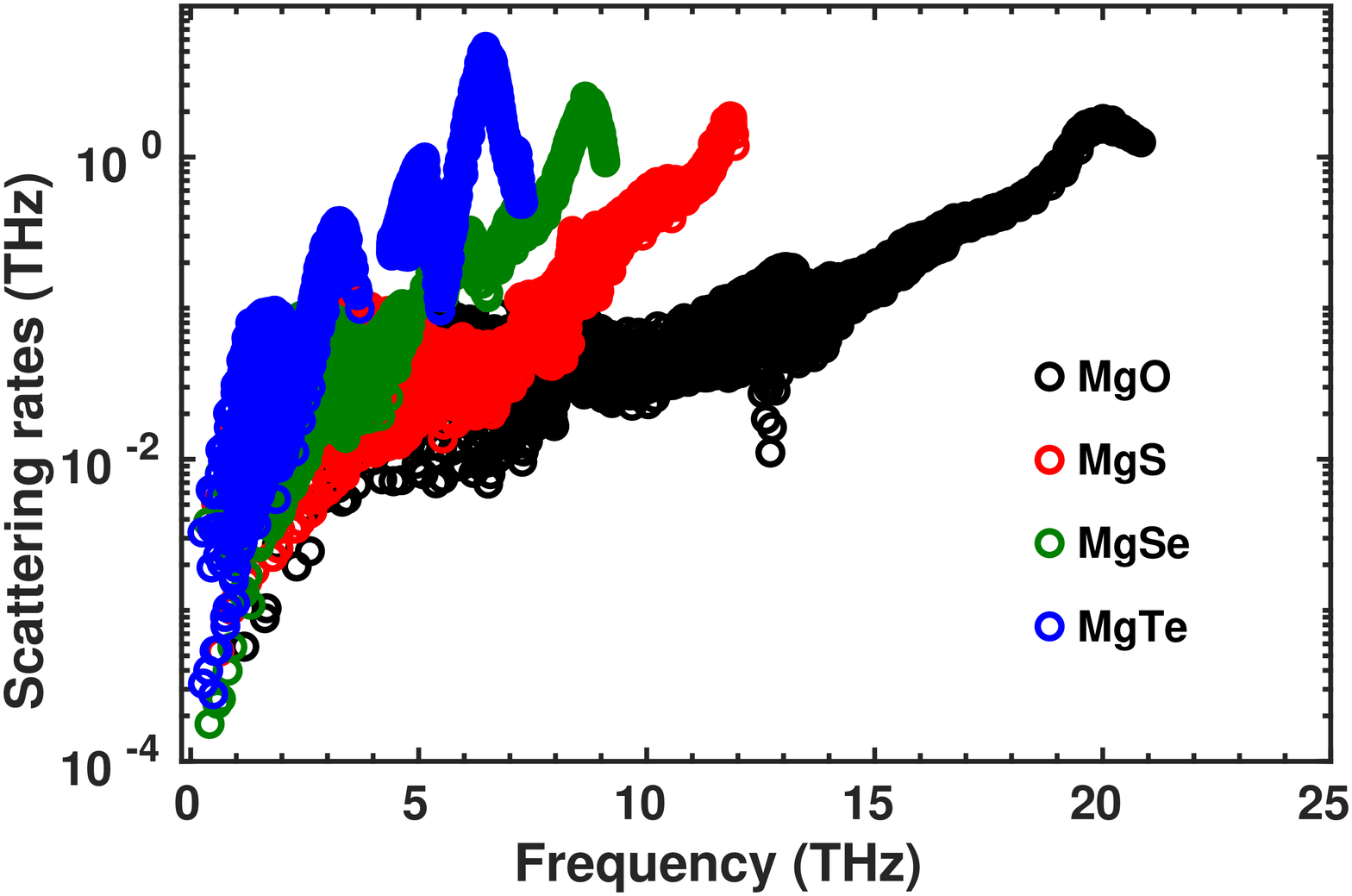}} 
\subfigure[]{\includegraphics[width=3.0in,height=2.0in]{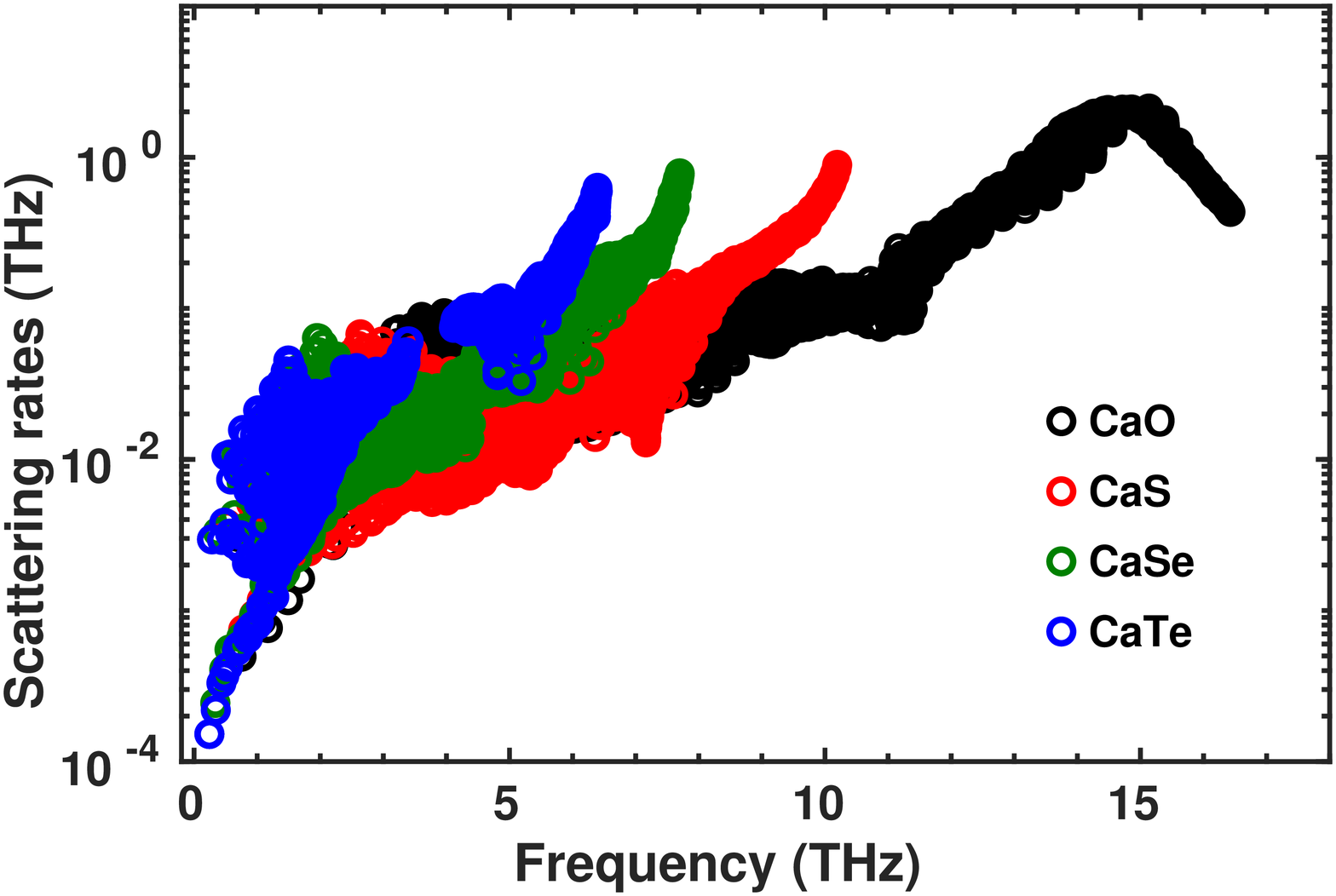}} \vspace{0.3in} \\
\subfigure[]{\includegraphics[width=3.0in,height=2.0in]{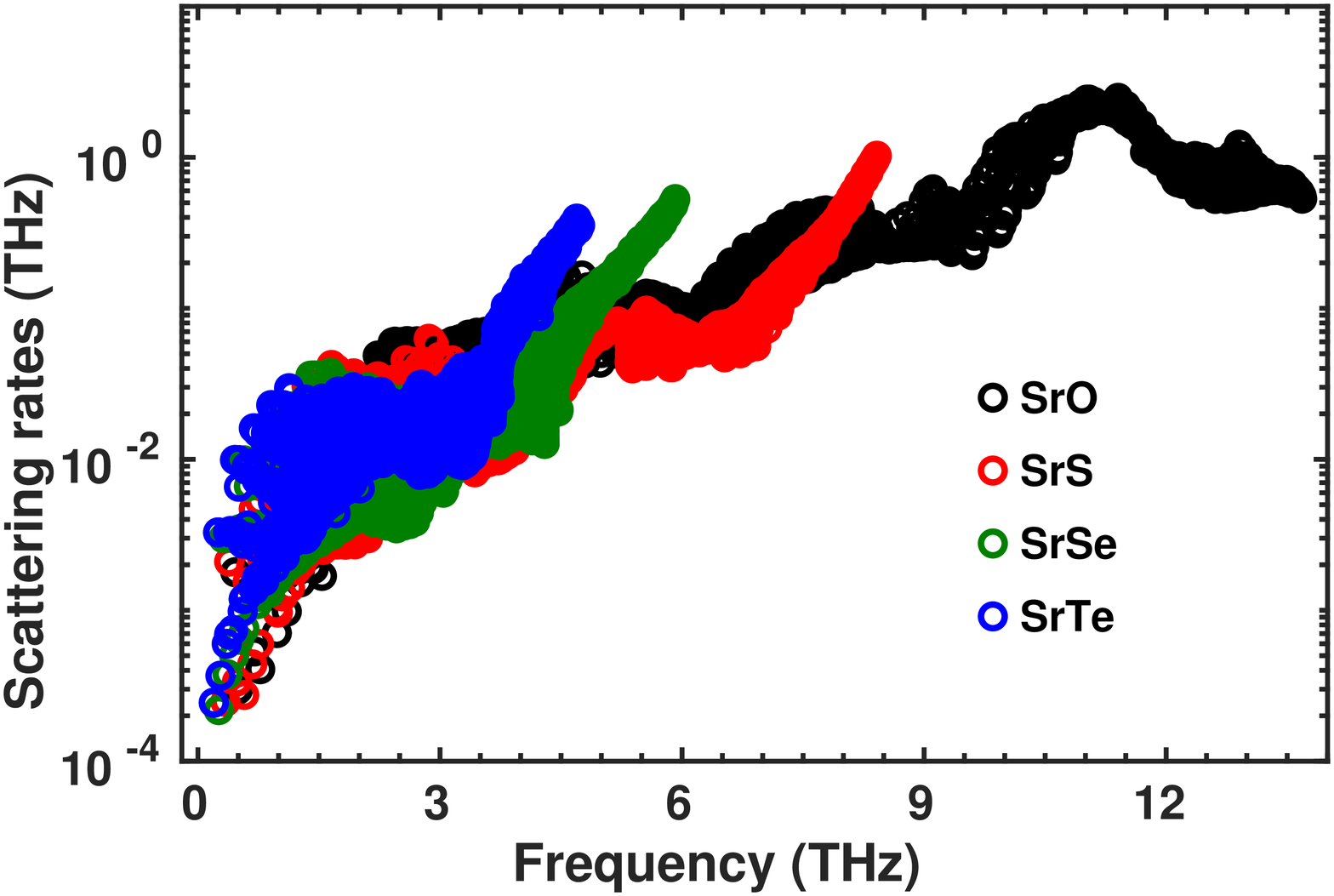}} 
\subfigure[]{\includegraphics[width=3.0in,height=2.0in]{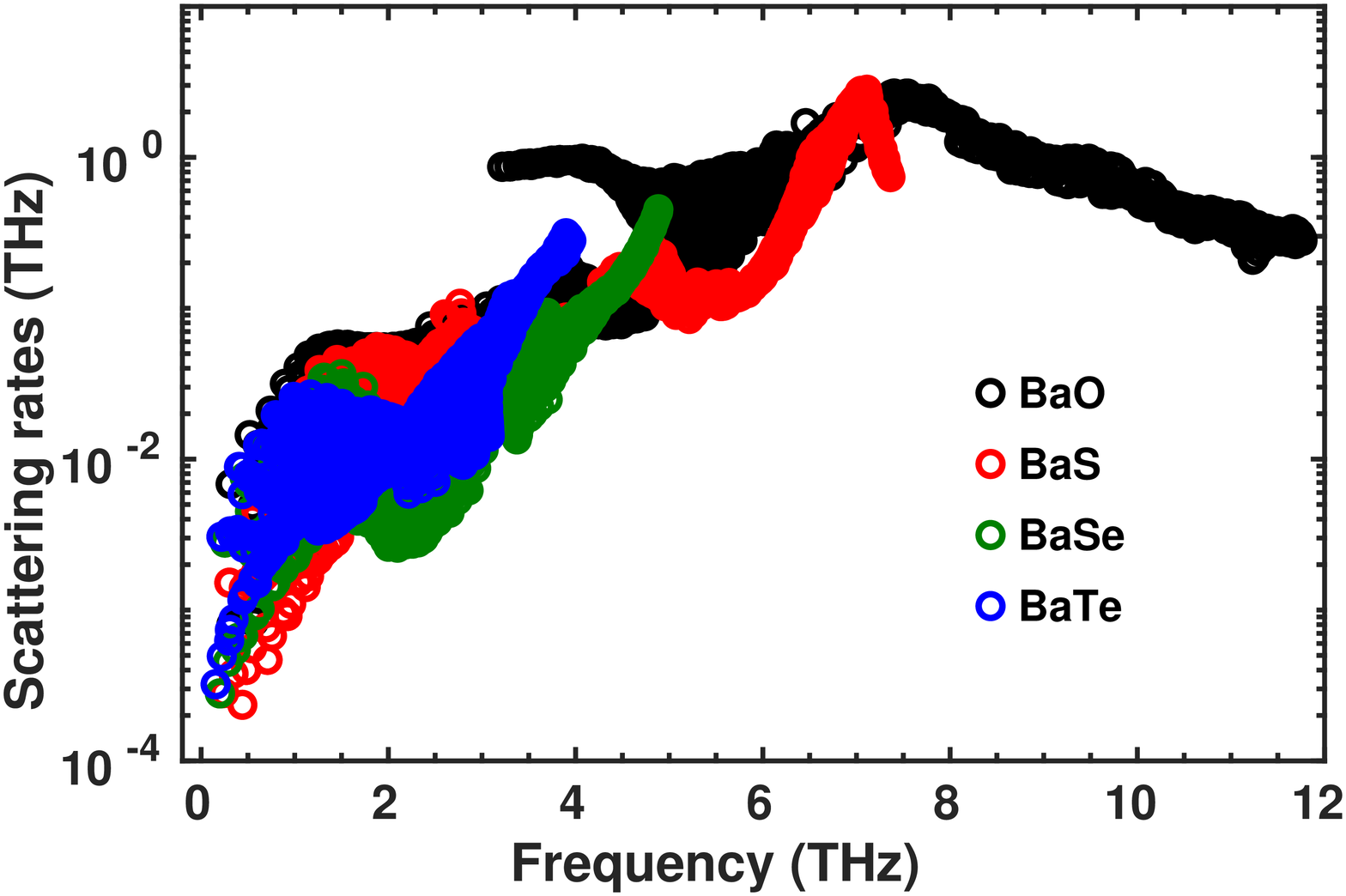}}
\caption{Calculated scattering rates of (a) MgX, (b) CaX, (c) SrX and (d) BaX compounds as a function of frequency; where X = O, S, Se and Te.}
\label{fig:SR}
\end{figure}

\begin{figure}
\centering
\subfigure[]{\includegraphics[width=3.0in,height=2.0in]{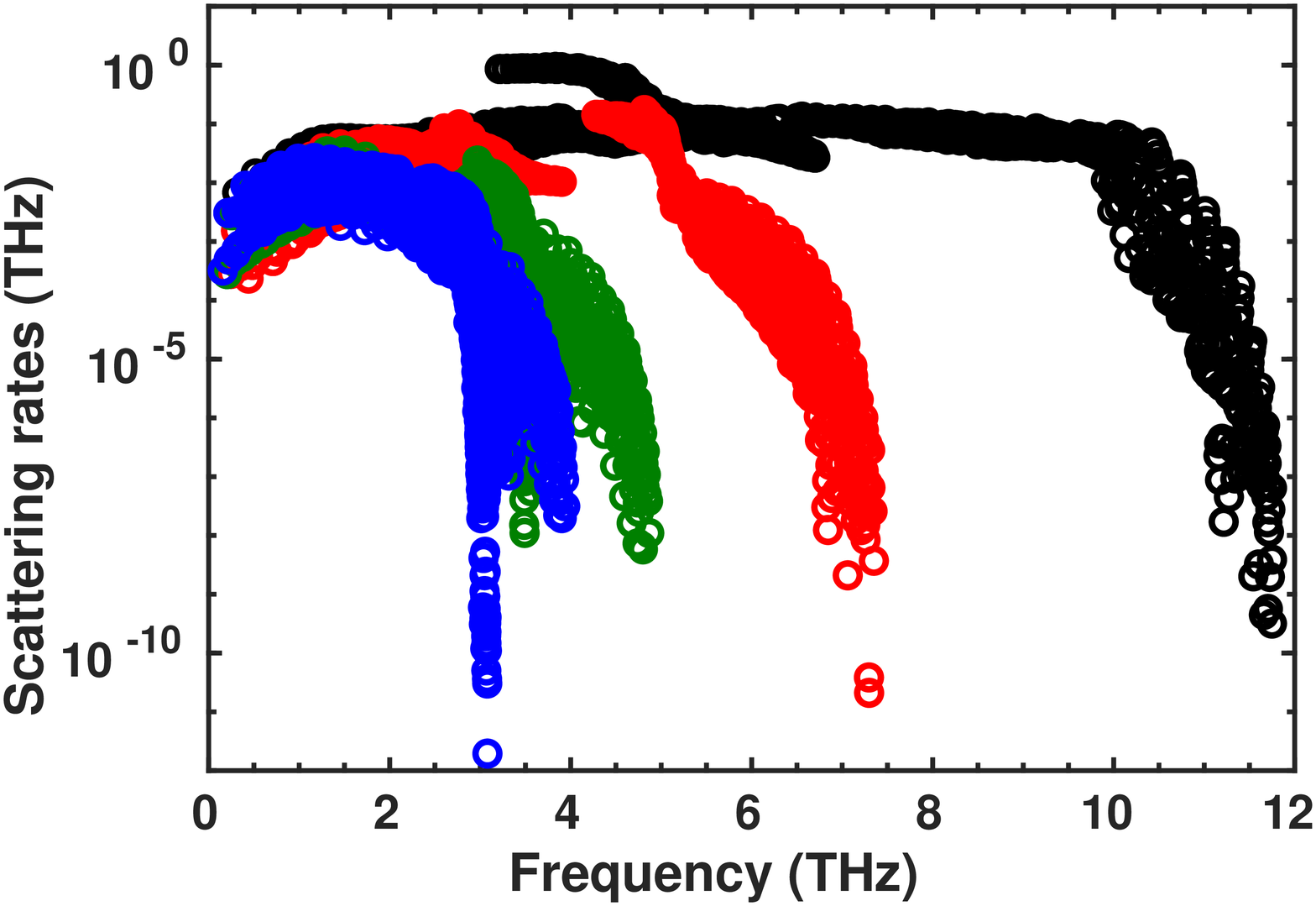}} 
\subfigure[]{\includegraphics[width=3.0in,height=2.0in]{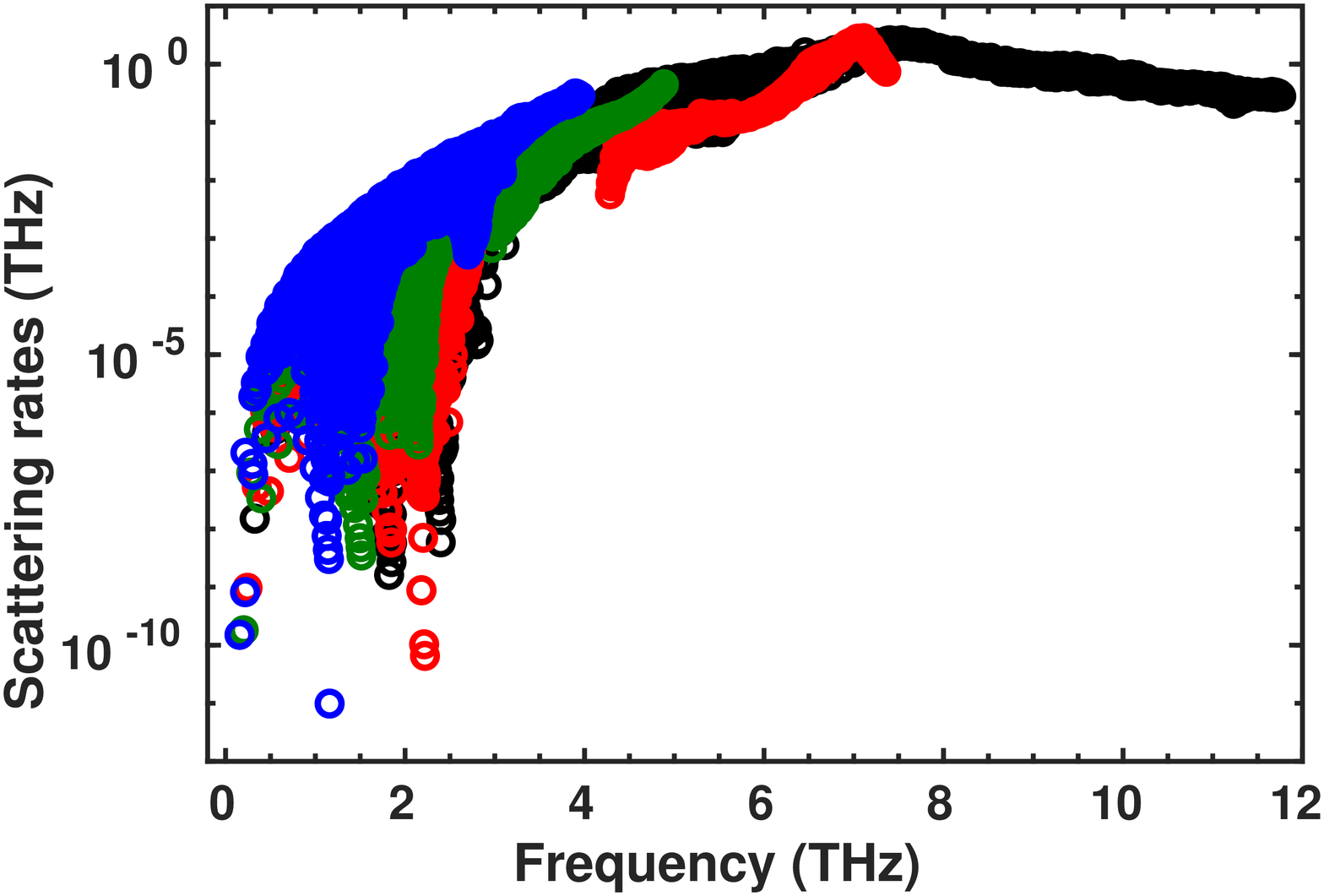}} \vspace{0.3in} \\
\subfigure[]{\includegraphics[width=3.0in,height=2.0in]{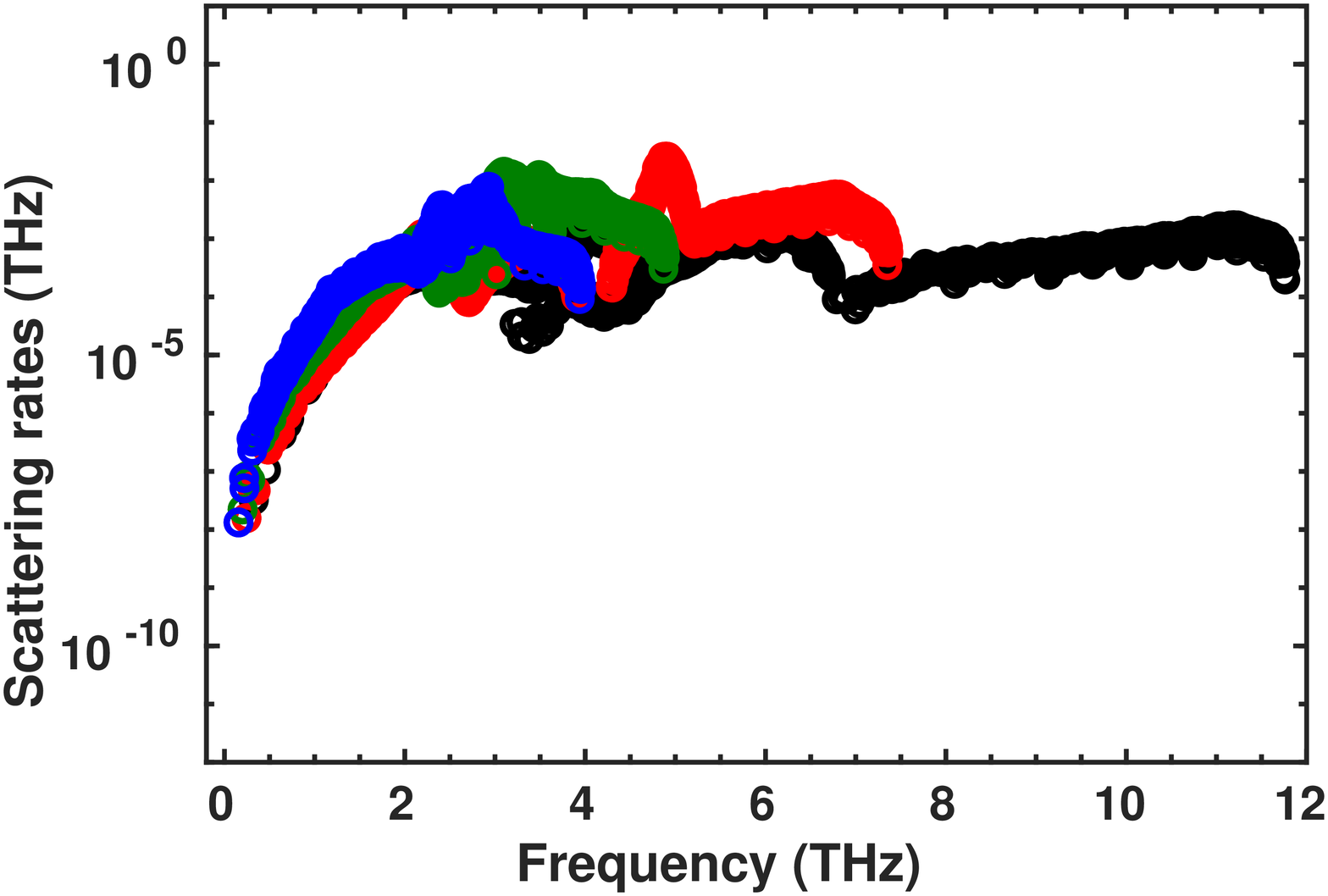}} 
\subfigure[]{\includegraphics[width=3.0in,height=2.0in]{Figures/BaX-SR.eps}}
\caption{Calculated (a) absorption (b) emission (c) isotope and (d) total scattering rates of BaX (X = O, S, Se and Te) compounds as a function of frequency.}
\label{fig:SR}
\end{figure}

\begin{figure}
\centering
\subfigure[]{\includegraphics[width=5.0in,height=3.5in]{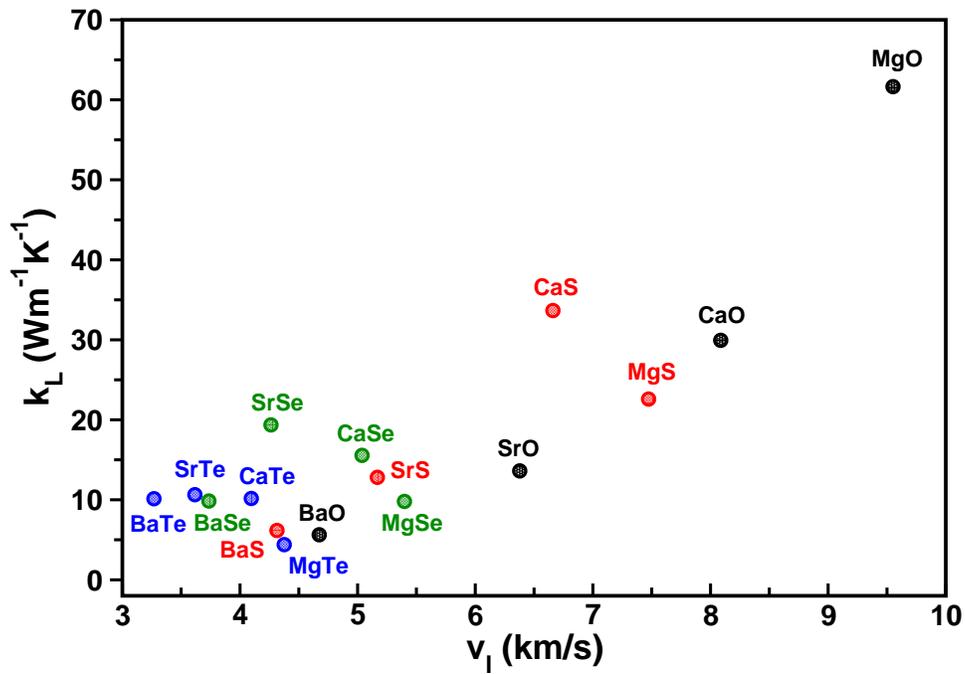}} \vspace{0.5in} \\
\subfigure[]{\includegraphics[width=5.0in,height=3.5in]{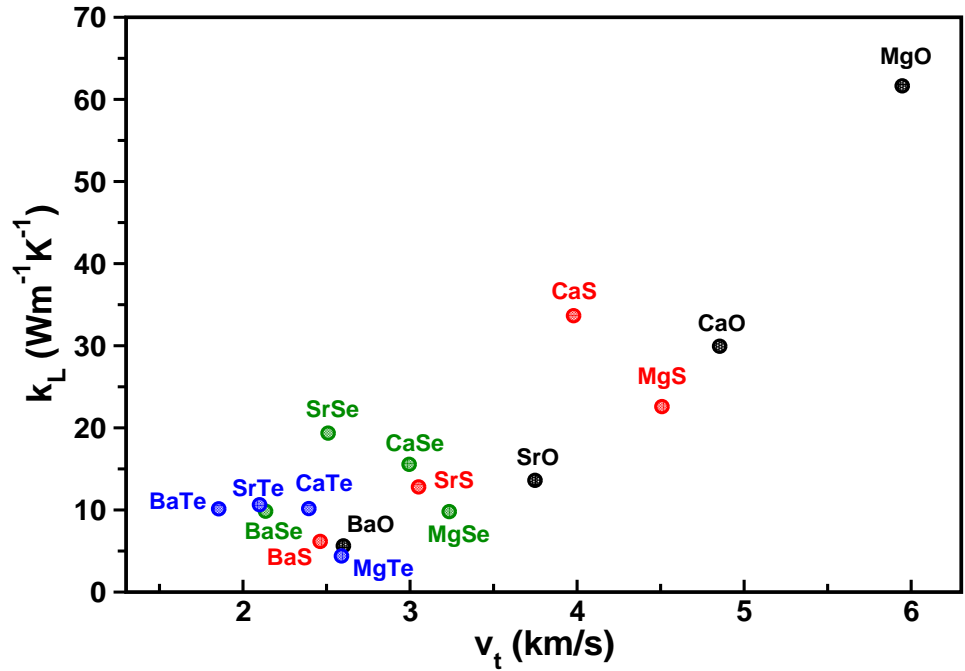}}
\caption{Calculated $k_L$ as a function of (a) longitudinal (v$_l$) (b) transverse (v$_t$) sound velocities for 16 MX compounds.}
\label{fig:Theta}
\end{figure}

\begin{figure}
\includegraphics[width=5.5in,height=3.4in]{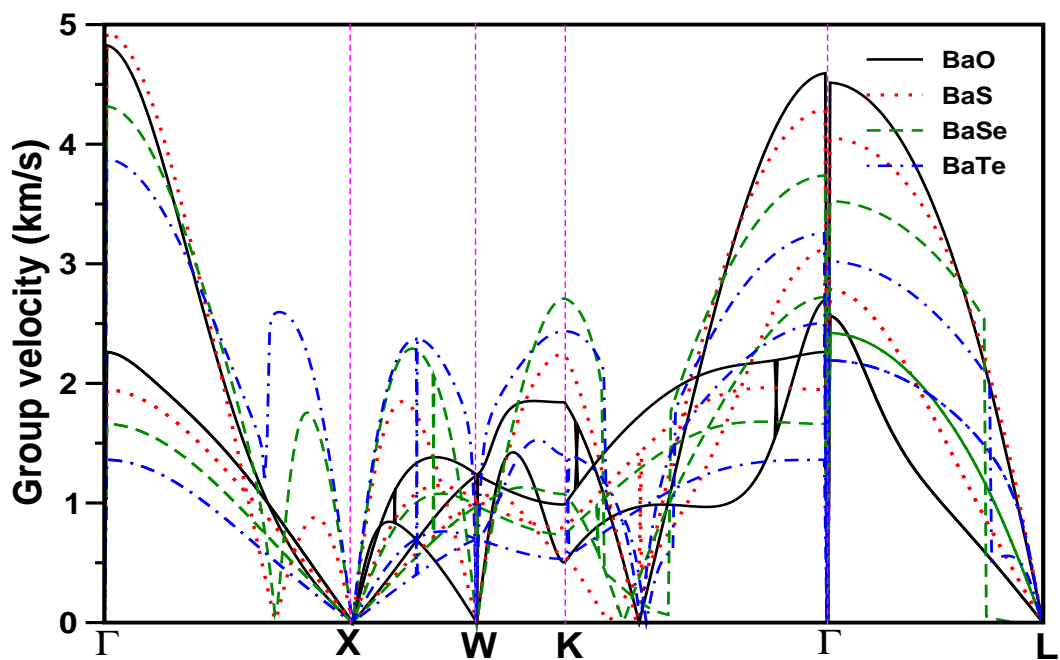}
\caption{Calculated phonon group velocities of BaX (X = O, S, Se and Te) compounds along high symmetry directions of the Brillouin zone.}
\label{fig:Theta}
\end{figure}


\clearpage

\begin{table}[tbp]
\caption{Calculated transverse optical ($\omega_{TO}$, in cm$^{-1}$), longitudinal optical ($\omega_{LO}$, in cm$^{-1}$) phonon modes and difference between LO and TO modes ($\omega_{LO}$-$\omega_{TO}$, in cm$^{-1}$) at 300 K for 16 MX compounds are compared with the available experimental and previous first principles calculations. The values given in parenthesis are in THz units.}
\label{table1}
\begin{tabular}{cccccc} \hline
Compound & Method  &  $\omega_{TO}$    &  $\omega_{LO}$     & $\omega_{LO}$-$\omega_{TO}$  \\ \hline
MgO 	& This work   &  390.11 (11.68)	& 696.06 (20.84) &  305.94 (9.16)   \\ 
MgS    & This work   &  243.82 (7.3) &	398.46 (11.93)  &	 154.64 (4.63) \\ 
& others  & 220.64$^a$ ,224.9$^b$ & 385.06$^a$,381.9$^b$ &165.06  \\
MgSe   	& This work &	 185.04 (5.54) &	 331.33 (9.92)&	 146.29 (4.38) \\ 
& others & 171$^c$,169.92$^a$ & 296$^c$, 292.31$^a$ &123.31 \\
MgTe    	& This work &	 147.29 (4.41)&	 243.49 (7.29) &  96.19 (2.88)  \\ 
& others & 136.38$^a$, 141$^d$& 236.06$^a$, 234.1$^d$ &99.68  \\
CaO  	& This work &	 317.63 (9.51) &	 588.51 (17.62)&	 270.87 (8.11)  \\ 
&Expt  & 311$^e$ & 585$^e$  & -  \\
CaS	 	& This work &  224.11 (6.71) & 340.35 	(10.19) &	 116.23 (3.48) \\ 
&others &  231.42$^b$  &336.65$^b$    &-       \\
&Expt &  229$^f$    & 342$^f$     &-        \\
CaSe  	& This work &	 177.02 (5.3)	 &  257.18 (7.7) &	 80.16 (2.4)  \\ 
&others & 178.4$^d$&  251.4$^d$ &- \\
&Expt & 168$^f$  &  220$^f$   &-  \\
CaTe  	& This work &	 155.98 (4.67) &	 213.76	 (6.4) & 57.78 (1.73) \\ 
& others & 156.2$^d$&  209$^d$ &- \\
\hline
\end{tabular}
\newline

$^a$Ref.\cite{MgX-Mir2016} $^b$ Ref.\cite{Ca-MgS-Ghebouli2014} $^c$ Ref.\cite{MgSeNaClWu2015} $^d$ Ref.\cite{Ca-Mg-Se-Te-Ghebouli2013} $^e$ Ref.\cite{MO-Expt-Galtier1972} $^f$ Ref.\cite{Ba-Sr-Ca-X-Kaneko1982}  $^g$ Ref.\cite{SrO-Souadkia2012} $^h$Ref.\cite{SrO-Expt-Rieder1975}  $^i$Ref. \cite{SrX-Phonons-Souadkia2014}
\end{table}

\begin{table}[tbp]
\caption*{Table S1 continued}
\label{table1}
\begin{tabular}{cccccc} \hline
Compound & Method  &  $\omega_{TO}$    &  $\omega_{LO}$     & $\omega_{LO}$-$\omega_{TO}$  \\ \hline
SrO   	& This work &	 202.74 (6.07)&	 457.25 (13.69) &	 254.51 (7.62)   \\ 
&others & 217.85$^g$& 474.13$^g$ & -  \\
&Expt & 224.15$^h$ & 484.99$^h$ & - \\
SrS  	& This work	&  179.02 (5.36)	&  280.90 (8.41) &	 101.87 (3.05) \\ 
&others & 194.045$^i$ & 290.577$^i$& -  \\
SrSe  	& This work &	 136.60 (4.03) & 197.73 	(5.92) &	 63.13 (1.89) \\ 
&others & 127.496$^i$ & 193.193$^i$ & - \\
SrTe  	& This work &	 113.90 (3.41) &	158.32 (4.74)  &	 44.42 (1.33)  \\ 
&others & 110.717$^i$ & 156.738$^i$  & -  \\
BaO	 	& This work &  107.55 (3.22)	&  392.78 (11.76)	 & 	285.24 (8.54)   \\ 
&Expt & 146$^e$ & 440$^e$  & -  \\
BaS 	& This work	&  142.95 (4.28)	& 245.82  (7.36)	  &  102.87 (3.08)	\\ 
&Expt & 150$^f$ & 246$^f$ & -  \\
BaSe 	& This work& 99.19 	(2.97) &	 162.99 (4.88) & 63.79 	(1.91)  \\ 
& Expt & 100$^f$ & 156$^f$  & - \\
BaTe 	& This work&	 90.18 (2.7)	&  131.93 (3.95)	 & 41.75  (1.25)	 \\ \hline
\end{tabular}
\newline
$^e$ Ref.\cite{MO-Expt-Galtier1972} $^f$ Ref.\cite{Ba-Sr-Ca-X-Kaneko1982}
\end{table}

\begin{table}[tbp]
\caption{Calculated second order elastic constants (C$_{11}$, C$_{12}$, C$_{44}$, in GPa) for MX compounds in rocksalt structure type.}
\label{table1}
\begin{tabular}{cccccccccc} \hline
         &            &  This work  &            &            & Others  &              \\ \hline    
Compound &  C$_{11}$  & C$_{12}$   &  C$_{44}$   &  C$_{11}$  & C$_{12}$   &  C$_{44}$   \\ \hline
MgO   &  298.03	 & 88.15 &	143.96 & 297 $^a$  &99.6 $^a$ &151.9$^a$     \\    
       &  296.47$^b$  & 95.069$^b$  & 155.89$^b$      &      -     & -   &  -   & \\
       &  297.9$^c$  & 95.8$^c$  & 154.4$^c$      &      -     & -   &  -   & \\
MgS    &  153.47	&39.75	&53.58   & 168.4 $^d$ & 42.2$^d$ &  55.2 $^d$     \\
MgSe   &  127.78	& 32.90	& 42.55  &112.66 $^e$ & 33.49 $^e$   & 42.65 $^e$      \\ 
MgTe   &  100.58 &	24.88 &	29.99  &94 $^f$  &24  $^f$ &29  $^f$       \\ 
CaO    &  230.72 &	58.90 &	77.14 &198.8  $^g$ &  57.1$^g$ &75.3$^g$   \\
       &      -     & -   &  81$^h$  &        -     & -   &  -   & \\
       &      220.53$^i$      & 57.67$^i$   &  80.03$^i$   &        -     & -   &  -   & \\
CaS    &  139.88 &	23.27&	34.05    &122.1 $^g$&23.9$^g$ &33.5 $^g$   \\
CaSe   &  119.12	&18.97	 & 27.42 &104.3 $^g$&19.4$^g$ &26.6$^g$      \\ 
CaTe   &  94.34	&13.92	 & 18.70   &94.13 $^j$   &13.76 $^j$ & 17.34 $^j$    \\ 
SrO    &  186.69	&48.25&	55.15  &159.7 $^k$  &46.7$^k$ &54.3$^k$    \\ 
      &      -     & -   &  56$^h$   &        -     & -   &  -   & \\
SrS    &  122.77	&18.40&	26.17  &109.7 $^k$&19.1$^k$ &26.3 $^k$ \\
SrSe   & 106.57	&14.63&	21.33      &93.6$^k$ &15.2$^k$ &21.1 $^k$\\ 
SrTe   &  86.14	&10.33&	14.75       & 54.8$^l$ &13$^l$ &21 $^l$ \\ 
BaO    & 142.02&	44.61&	36.04    &121.99 $^m$ &   42.12 $^m$ &36.33 $^m$ \\ 
       &      -     & -   &  34$^n$   &        -     & -   &  -   &  \\
BaS    & 104.74&	16.60&	18.58  &91.29 $^m$ &16.7 $^m$ & 18.74  $^m$\\
BaSe   & 93.43&	12.91&	15.53      &  78.42 $^m$ &13.14 $^m$ &15.61  $^m$ \\ 
BaTe   &  78.04&	8.56&	11.09      &  68.36 $^m$ &9.1 $^m$ &11.26 $^m$ \\ 
PbTe    & 126.49 & 4.31 & 14.88  & 136.2 $^o$ ,102.93 $^p$ &3.79 $^o$, 10.125$^p$  & 17.14 $^o$, 12.641$^p$  \\
        & 128.1$^q$ &4.4$^q$  & 15.14$^q$  &- &- &-   \\
\hline
\end{tabular}
\newline
$^a$Ref. \cite{MO-Elastic-CINTHIA201523} $^b$Ref.\cite{MgO-Elastic0ExptANDERSON1966} $^c$Ref.\cite{MgO-Expt-2-Sinogeikin1999} $^i$Ref.\cite{CaO-Elastic-expt-Oda1992} $^d$Ref.\cite{MgS-Se-Saib2009} $^e$Ref.\cite{MgSeNaClWu2015}   $^f$ Ref.\cite{MTeRen2017} $^g$Ref. \cite{Elastic-1}  $^h$Ref.\cite{CaSrO-Expt-elasticSON1972}  $^j$ Ref. \cite{CaX-RafikMaizi2019} $^k$ Ref.\cite{SrXAbdusSalam2019}  $^l$Ref.\cite{Thermodynamic-1} $^m$Ref.\cite{Rajput2021}  $^n$Ref.\cite{BaO-elastic-Vetter1973} $^p$Ref.\cite{PbTe-a-elastic-Xue2021} $^o$Ref.\cite{elastic-PbTe-Yang2012} $^q$Ref.\cite{PbTe-Elastic_C_Dornhaus1983}
\end{table}

\begin{table}[tbp]
\caption{Calculated Young's modulus (E, in GPa), Bulk modulus (B, in GPa), Shear modulus (G, in GPa),  density ($\rho$, in gr/cc),  sound velocities (v$_l$, v$_t$ and v$_m$, in km/s), Debye temperature ($\Theta_D$, in K), Poisson's ratio ($\sigma$) and Gr\"uneisen parameter  ($\gamma$) for MX compounds.}
\label{table1}
\begin{tabular}{ccccccccccccccccccc} \hline
Compound    & E &B & G     & $\rho$   &$v_l$     & $v_t$  & $v_m$   & $\Theta_D$  & $\sigma$ & $\gamma$     & \\ \hline
MgO      &300.23 &158.11 &126.84  &3.59&	9.55&	5.95	&6.55	&927.14 &0.18   &1.23    \\    
Others  &305$^a$  &165.5$^a$ & 127.9$^a$&3.0$^a$&9.826$^a$  &6.115$^a$  &6.740$^a$  &902$^a$ &0.252$^a$  &-\\
 
MgS      & 133.23& 77.66& 54.87  &  2.70&	7.47&	4.51&	4.98	&573.55   &0.21   &1.24     \\
Others  &- &73.102$^b$ &-&- &- &- &- &- &- &- \\
   & &79.44$^c$ &     &  	&	& &      	&	&   &  &  \\ 
   
   & &72.3$^d$ &     &  	&	& &      	&	&   &  &  \\ 
MgSe        & 108.43& 64.52&44.44   &  4.25&	5.40&	3.23&	3.58	&391.45  &0.22 &	1.36      \\
Others  &113.07$^e$ & 60.203$^b$& 46.28$^e$&- &- &- &- &- &0.2216$^e$ &- \\
  & 109.2$^f$  & 67.7$^e$     &  45.7$^f$   & & & & & & & \\
     &   & 65.15$^c$      && & & & & & & \\

MgTe        &81.03 &50.11 &32.92   &  4.91&	4.38&	2.59&	2.87	&289.54   &0.23 &	1.41  \\ 
Others  &77$^g$&45.795$^b$ &31$^g$&- &4.420$^g$ &2.243$^g$ &- &251$^g$ &0.33$^g$&- \\

  & & 50.45$^c$      && & & & & & & \\
   & & 44.6$^d$     && & & & & & & \\
  
CaO   & 196.26&116.17&80.54    & 3.42	&8.09	&4.85	&5.37	&669.61    & 0.22 &	1.36  \\ 
Others &178.51$^h$ &104.32$^h$ &73.47$^h$ &3.2$^a$ &8.394$^a$ &5.085$^a$ &5.619$^a$ &691$^a$  &0.21$^h$ &- \\

 & 205.6$^a$ & 118.8$^a$      & 84.88$^a$& & & & & & 0.221$^a$& \\
 
CaS   & 103.43&62.14&42.30   &  2.67	&6.66	&3.98	&4.40	&465.12    &0.22 &	1.37   \\
Others &95.27$^h$&56.62$^h$ &39.06$^h$ &- &- &- &- &- &0.22$^h$ &- \\
 & & 62.90$^i$    && & & & & & & \\
  & & 64$^j$    && & & & & & & \\

CaSe    & 85.82&52.35 &34.98  & 3.90&	5.04&	2.99&	3.32&	336.21      & 0.23	&1.39 \\ 
Others &78.69$^h$ &47.70$^h$ &32.12$^h$ &- &- &- &- &- &0.23$^h$ &- \\
& & 52.17$^i$    && & & & & & & \\
& & 51$^j$  && & & & & & & \\


\hline
\end{tabular}
\newline

$^a$Ref.\cite{MO-Elastic-CINTHIA201523} $^b$Ref.\cite{MgX-Theory-Tairi2017} $^c$Ref. \cite{Mg-CaX-Debnath2018} $^d$Ref.\cite{MgX-Mir2016} $^e$Ref.\cite{MgSe-Muthaiah2021} $^f$Ref.\cite{MgSeNaClWu2015} $^g$ \cite{MTeRen2017} $^h$Ref.\cite{Elastic-1} $^i$Ref.\cite{CaX-RafikMaizi2019} $^j$Ref. \cite{Luo1994} $^k$Ref.\cite{Charifi2005}
\end{table}

\begin{table}[tbp]
\caption*{Table S3 continued}
\label{table1}
\begin{tabular}{cccccccccccccccccccc} \hline
Compound    & E &B & G     & $\rho$   &$v_l$     & $v_t$  & $v_m$   & $\Theta_D$  & $\sigma$ & $\gamma$     & \\ \hline
CaTe    & 63.39&40.73&25.55   &  4.46&	4.10&	2.40&	2.65 &	251.08  & 0.24 &	1.45 \\ 
Others&62$^g$ &39.6$^k$ &25$^g$  &- &4.246$^g$ &1.998$^g$  &- &209$^g$  &- &-  \\
& & 40.45$^i$ && & & & & & & \\
& & 41.8$^j$  && & & & & & & \\
SrO    &149.36 &94.4 &60.41    &  4.30&	6.38&	3.75	&4.15	&481.98 & 0.24 &1.43      \\   
Others &135.81$^l$ &82.4$^l$  &55.13$^l$  &4.9$^a$&5.743$^a$ &3.385$^a$ &3.751$^a$ &430$^a$ &0.23$^l$  &- \\

& 139 $^a$ & 87.6$^a$ & 56.28$^a$& & & & & & 0.222$^a$& \\

SrS     &85.36 &53.19 &34.63    &  3.72&	5.17&	3.05&	3.38	&336.80  &0.23 &	1.42  \\
Others &80.48$^l$  &47.3$^l$  &32.77$^l$  &- &- &- &- &- &0.23$^l$  &- \\

SrSe     &72.04 &45.28 &29.17   &  4.63&	4.26&	2.51	&2.78	&266.98   &0.23 &	1.43    \\ 
Others  &66.80$^l$  &40.3$^l$  &27.14$^l$  &- &- &- &- &- &0.23$^l$  &- \\

SrTe      &54.24 &35.6&21.77   &  4.94	&3.62 &	2.10 &	2.33	&209.78    &0.25 &1.48     \\ 
Others   &64.77$^m$ &22.12$^m$&23.54 $^m$&- &3.996$^g$ &1.714$^g$ &- &119.450$^m$ &0.019$^m$  &- \\

& & 39.5$^n$ && & & & &172$^g$ &0.39$^g$& \\

BaO  & 103.75 & 77.08 & 40.66 & 6.01 & 4.67 & 2.60 & 2.90 & 311.79 & 0.28 & 1.63 \\ 
Others  &107.45$^o$ &76.3$^a$ &41.88$^a$ &5.7$^a$&4.76$^a$ &2.672$^a$ &2.973$^a$&316$^a$ &0.35$^o$ &- \\

 & 106.2$^a$ &  && & & & & & 0.255$^a$ & \\

BaS   &66.63 &45.98 &26.47   & 4.37	&4.31	&2.46	&2.74	&256.14      &0.26 &1.54 \\
Others &88.24$^o$ & 41.6$^p$ &24.935$^p$&- &4.2$^q$ &2.37$^q$  &2.67$^p$ &247.26$^p$  &0.18$^o$ &- \\

& &  && & & &2.64$^q$ &247$^q$ &0.25$^p$ & \\

BaSe    &57.85 &39.75 &23.00 &  5.05	& 3.73 &	2.13	& 2.38 &	214.81 &0.26 &1.53     \\ 
Others  &76.21$^o$, &34.0$^r$ &21.41$^p$ &- &3.64$^q$ &2.07$^q$  & 2.3$^p$  & 205.7$^p$ &0.17$^o$ &- \\

& & 36.267$^p$  && & & &2.31$^q$&208$^q$  &0.252$^s$ & \\
& &  && & & & & & 0.25$^p$ & \\

BaTe    &45.17 &31.72 &17.89     &  5.20	& 3.27 &	1.85 &	2.06&	176.28  & 0.26  &1.56    \\
Others  &67.15$^o$ &27.04$^r$ &16.697$^p$&- &3.68$^g$ &1.46$^g$ &1.99$^p$ &171.22$^p$  &0.13$^o$&- \\

& & 28.23$^p$ && & & & &139$^g$ & 0.271$^s$ & \\
& &  && & & & & & 0.25$^p$ & \\
PbTe    & 68.24 & 45.04 & 27.35 & 8.32 & 3.13 &1.81 &2.01 & 186.07 & 0.25 & 1.50 \\
Others &54.72$^t$ &41.06$^t$ &21.99$^t$&- &2.94$^t$ &1.64$^t$ &1.83$^t$  &266.51$^t$  &0.273$^t$ &- \\
\hline
\end{tabular}
\newline
$^a$Ref.\cite{MO-Elastic-CINTHIA201523} $^g$ Ref.\cite{MTeRen2017} $^l$Ref.\cite{SrXAbdusSalam2019} $^m$Ref.\cite{SrTe-lattice-Saoud2016} $^n$ Ref.\cite{Zimmer1985} $^o$ Ref. \cite{Rajput2021} $^p$Ref.\cite{Thermo-2} $^q$Ref.\cite{BaX-Gkolu2008} $^r$Ref.\cite{BaSe-Te-Drablia2017} $^s$Ref.\cite{Khalfallah2018} $^t$Ref.\cite{PbTe-a-elastic-Xue2021} 
\end{table}

\clearpage
\bibliography{Refs.bib}